\documentclass[graybox, secnum]{svmult}

\usepackage{mathptmx}       
\usepackage{helvet}         
\usepackage{courier}        
\usepackage{type1cm}        
\usepackage{makeidx}         
\usepackage{graphicx}        
\usepackage{multicol}        
\usepackage[bottom]{footmisc}
\usepackage{hyperref}        
\usepackage{soul}            
\hypersetup{colorlinks=true,urlcolor=blue}
\usepackage[square,numbers]{natbib}
\makeindex             
\usepackage{amssymb}
\usepackage[mathscr]{euscript}
\usepackage{aas_macros}

\newcommand{\VB}[1]{\textcolor{black}{#1}}

\newcommand{\FM}[1]{\textcolor{black}{#1}}
\newcommand{\FMbis}[1]{\textcolor{black}{#1}}
\newcommand{\VBbis}[1]{\textcolor{black}{#1}}

\begin{document}
\title*{Chemical enrichment in groups and clusters}
\author{Fran\c{c}ois Mernier\thanks{corresponding author}\thanks{\textit{ESA Research Fellow.}} and Veronica Biffi}
\institute{Fran\c{c}ois Mernier \at European Space Agency (ESA), European Space Research and Technology Centre (ESTEC), Keplerlaan 1, 2201 AZ Noordwijk, The Netherlands.\\ \email{francois.mernier@esa.int}
\and
Veronica Biffi \at 
INAF - Osservatorio Astronomico di Trieste, via Tiepolo 11, I-34143 Trieste, Italy\\
\email{veronica.biffi@inaf.it}
}
%
%
\maketitle

\abstract{As building blocks of dust, rocky planets, and even complex life, 
\VB{the chemical elements heavier than hydrogen (H) and helium (He) -- called "metals" in astronomy --}
play an essential role in our Universe and its evolution.  
\VB{\FM{Up to Fe \FMbis{and Ni},} these metals}
are known to be created by stars and stellar remnants \FM{via} nuclear fusion and to be ejected into their immediate surroundings to enrich new stellar generations. A spectacular finding, however, is that these processed elements are found even \textit{outside} stellar systems, in particular in the hot, X-ray atmospheres surrounding early-type galaxies and pervading galaxy clusters and groups. These large-scales structures are thus a remarkable fossil record of the integrated history of the enrichment of our Universe. 
In this Chapter, we briefly discuss 
\VB{the chemical properties of}
this intracluster -- or intragroup -- medium (ICM). 
After  
\VB{introducing} the concept of chemical abundances, 
\VB{and recalling}
which stellar sources produce which elements, 
\VB{we review the method to derive abundance measurements}
from observations of the ICM and  
\VB{detail some chemical models implemented in}
numerical hydrodynamical simulations of cosmic structures. 
In particular, we explore how synergies between X-ray observations and numerical simulations help us to understand (i) the cosmic epoch at which the bulk of the enrichment occurred, (ii) the physics of these stellar sources responsible for this enrichment, and (iii) the main mechanisms responsible for metal diffusion and mixing outside galaxies.}
\section{Keywords} 
Intracluster medium -- X-ray -- Metals -- Chemical enrichment -- Galaxy clusters -- Galaxy groups -- Early-type galaxies -- Numerical simulations -- Spectroscopy

\section{Introduction}
\label{sec:intro}

From the cosmological paradigm currently favoured by scientists -- the $\Lambda$CDM model \FM{(see Chapter: X-ray cluster cosmology)}, \FMbis{gravity as a fundamental force has built the large scale structures of our Universe in a hierarchical way}. Dark matter, \FM{\FMbis{whose} exact nature is still to be unveiled,} is sensitive to nothing else than gravity\FMbis{. It has} progressively assembled into a network of filaments, then \FMbis{has been} forming and falling into deep gravitational wells. Baryons, a.k.a. the ``ordinary'' (or ``visible'') matter, have been naturally guided through these gigantic flows during their own collapse \FM{(see Chapter: Cluster outskirts and their connection to the cosmic web)}. But on the contrary of dark matter, the sensitivity of baryons to the other fundamental forces (electromagnetic, nuclear weak and strong) made them able to emit, absorb, and interact with light. Baryons took then advantage of these pre-shaped gravitational wells and valleys to assemble rapidly and eventually partly cool down into stars, atomic and molecular gas, dust, and planets. Situated at the knots of this large-scale network, galaxy groups and clusters are thus made of galaxies, but not only. In fact, about $\sim$80\% of the visible matter in these systems is made of a plasma, the intracluster medium (ICM)\footnote{When referring to galaxy groups, this hot atmosphere is often mentioned in the literature as the intragroup medium (IGrM). For simplicity, in this Chapter we will refer to the ICM regardless of 
\VB{the size of the host (from groups of few galaxies to clusters of several tens\FMbis{, hundreds, or even thousands} of galaxies)}.} \FM{(see Chapter: Thermodynamical profiles of clusters and groups, and their evolution)}, while stars in galaxies account only for the remaining 20\% \citep{giodini2009}. Having been heated by shocks and compressed adiabatically during its collapse, this gas is in fact extremely hot ($10^{7-8}$~keV), which makes it shine in the X-ray band of the electromagnetic spectrum. This means that flying X-ray observatories (\textit{XMM-Newton}, \textit{Chandra}, etc.) are extremely useful to detect the ICM in clusters and groups, as well as to investigate its physical properties -- essentially via X-ray spectroscopy.

As \textit{the} major baryonic component of clusters and groups, understanding the assembly and dynamics of the ICM, as well as possible feedback mechanisms 
driving its interplay with the galaxy population \FM{(see Chapter: AGN feedback in groups and clusters)},
is absolutely essential to fully picture our Universe at its largest scales. These major topics are addressed \FM{by other Chapters of this section} \citep[for another comprehensive review, see also][]{werner2020}.

Another, no less interesting aspect of the ICM is its chemical composition. As this gas was once cold and widely dispersed before collapsing onto clusters, one might easily suppose that today's ICM should reflect the composition of the primordial nucleosynthesis (i.e. $\sim$75\% of H and $\sim$25\% of He). Yet, the first flying X-ray spectrometers in the late 70's revealed the surprising presence of metals\footnote{In astronomy, ``metals'' traditionally refer to all chemical elements heavier than He.} in this hot gas \citep{mitchell1976}. This finding has in fact profound consequences: since the only place where metals can be produced is the burning cores of stars and supernovae, this gas must have been enriched by stellar populations from galaxies at some stage of its evolution. Not only this reveals a remarkable interplay between stars, galaxies, and the ICM, but 
also brings the evidence that the elemental bricks of dust, asteroids, rocky planets (and even life!) are found out to the largest scales of the Universe. Understanding the full cycle of metals in galaxies and clusters necessarily goes through answering two fundamental questions:

\begin{enumerate}
    \item \textbf{How} and \textbf{when} did the 
    \FM{large scale hot gas}
    become chemically enriched?
    \item Can metals in the \FM{large scale hot gas} help us to better understand \textbf{stellar physics}?
\end{enumerate}

These questions are of course very complex and require a tight  
\VB{synergetic investigation}
between (i) X-ray observations of clusters/groups -- to measure integrated and spatial distribution of metals in the ICM
and (ii) numerical simulations of cosmic structures -- to study the ICM and its evolution, with the aim of interpreting the observational findings and making further predictions. Generally speaking, observations can provide useful constraints to future simulations, and vice-versa. \FM{The} main results from these two approaches are further discussed throughout this Chapter.


\section{Abundances and metallicity}
\label{sec:formalism}
Before exploring further the concept of chemical enrichment, let us  
\VB{introduce} some formalism 
\VB{usually adopted to quantify the metal content}.

In observations of various astrophysical systems (stars, planets, galaxies, etc.), the amount of metals is measured based either on the global metal content of that system, or on one specific element. The former quantity is named \textbf{metallicity}, while the latter is named \textbf{(chemical) abundance}. 

\begin{enumerate}
    \item The \textbf{metallicity} $Z$ corresponds to the (mass) fraction of all metals in that given system. It is defined as 
    \begin{equation}\label{eq:metallicity}
        Z = 1 - X - Y,
    \end{equation}
    where $X$ and $Y$ are the mass fractions of H and He, respectively. For instance, in our Sun \VB{(depicted with the symbol $\odot$)}, one typically has $X_\odot \simeq $ \FMbis{0.73--0.74}, $Y_\odot \simeq 0.25$, and thus $Z_\odot \simeq $ \FMbis{0.01--0.02} \citep[e.g.][]{asplund2009}.
    \item The \textbf{abundance} of a given chemical element X, is usually a logarithmic function of the fraction of X atoms over hydrogen atoms contained in that system, with respect to that of the Sun. One usually writes
    \begin{eqnarray}\label{eq:abundance}
        \mathrm{[X/H]} &=& \log_{10} \left( \frac{N_\mathrm{X} / N_\mathrm{H}}{N_{\mathrm{X},\odot} / N_{\mathrm{H},\odot}} \right)\\
        &=& \log_{10} \left( \frac{N_\mathrm{X}}{N_\mathrm{H}} \right) - \log_{10} \left( \frac{N_{\mathrm{X},\odot}}{N_{\mathrm{H},\odot}} \right),
    \end{eqnarray}
    where $N_\mathrm{X}$ and $N_\mathrm{H}$ are respectively the number of atoms of element X and of hydrogen in the considered system, while $N_{\mathrm{X},\odot}$ and $N_{\mathrm{H},\odot}$ are the numbers of these corresponding atoms in our Sun, taken as reference. 
\end{enumerate}

The reference solar abundances are not known with infinite accuracy. In fact, various independent research groups have attempted to measure the abundance of our Sun from either spectra of its photosphere (``solar'' abundances) or directly from the composition of meteorites (``proto-solar'' abundances). While the former rather represents the abundances at the surface of today's Sun (which might not be necessarily reflecting those in its core), the latter is more representative of the Solar System at its very early stages of formation. The most widely used reference solar abundances in the literature are those from \citet[][hereafter AG89]{anders1989}. Admittedly, however, such measurements have been significantly revised since 
\VB{they were proposed}. In chronological order, other
\VB{proposed (proto) solar references that are currently widely used are:}
 \citet[][hereafter GS98]{grevesse1998}, \citet[][hereafter L03]{lodders2003}, \citet[][hereafter L09]{lodders2009}, and \citet[][hereafter A09]{asplund2009}. As shown in Table~\ref{tab:ref_abun}, these reference measurements agree well for some specific elements (e.g. Ca). Other key elements, however,  
 \VB{are characterised by} large differences (e.g. O, Ne, Fe). For this reason, it is very important to  
 \VB{always explicitely state}
 the (proto-) solar table used as reference when one  
 \VB{reports}
 observed or simulated abundances. This requirement is valid for this Chapter as well: throughout the text and figures we will systematically mention the reference units associated to each reported abundance value. 

\begin{table}
\begin{centering}
\caption{Reference abundance tables for chemical elements of interest of this Chapter, reported in units of atom number fractions relative to \FM{hydrogen}. \FM{Elements considered in this table (and, more generally, in this chapter) are: hydrogen (H), helium (He), carbon (C), nitrogen (N), oxygen (O), neon (Ne), magnesium (Mg), silicon (Si), sulfur (S), argon (Ar), calcium (Ca), chromium (Cr), manganese (Mn), iron (Fe), and nickel (Ni).} The $\Delta_\mathrm{max}$ value represents the largest relative difference (\FM{$|1-\mathrm{max}/\mathrm{min}|$, in percent}) between two reference abundance measurements for a given element.}             
\label{tab:ref_abun}
\setlength{\tabcolsep}{7pt}
\begin{tabular}{lcccccc}        
\hline \hline                
El. & AG89 & GS98 & L03 & L09 & A09 & $\Delta_\mathrm{max}$  \\   
 & (solar) & (solar) & (solar) & (proto-solar) & (solar) & (\%)  \\   
 & \citep{anders1989} & \citep{grevesse1998} & \citep{lodders2003} & \citep{lodders2009} & \citep{asplund2009} &  \\   
\hline
H		&	1.0	&	1.0	&	1.0	&	1.0	&	1.0 & $-$	\\
\FM{He}		&	$9.77 \times 10^{-2}$	&	$8.51 \times 10^{-2}$	&	$7.92 \times 10^{-2}$	&   $9.69 \times 10^{-2}$	&	$8.51 \times 10^{-2}$	&	23	\\
\FM{C}		&	$3.63 \times 10^{-4}$	&	$3.31 \times 10^{-4}$	&	$2.45 \times 10^{-4}$	&   $2.78 \times 10^{-4}$	&	$2.69 \times 10^{-4}$	&	48	\\
\FM{N}		&	$1.12 \times 10^{-4}$	&	$0.83 \times 10^{-4}$	&	$0.68 \times 10^{-4}$	&   $0.82 \times 10^{-4}$	&	$0.68 \times 10^{-4}$	&	65	\\
O		&	$8.51 \times 10^{-4}$	&	$6.76 \times 10^{-4}$	&	$4.90 \times 10^{-4}$	&   $6.05 \times 10^{-4}$	&	$4.90 \times 10^{-4}$	&	74	\\
Ne		&	$1.23 \times 10^{-4}$	&	$1.20 \times 10^{-4}$	&	$0.74 \times 10^{-4}$	&	$1.27 \times 10^{-4}$	&	$0.85 \times 10^{-4}$	&	72	\\
Mg		&	$3.80 \times 10^{-5}$	&	$3.80 \times 10^{-5}$	&	$3.55 \times 10^{-5}$	&	$3.97 \times 10^{-5}$	&	$3.98 \times 10^{-5}$	&	12	\\
Si		&	$3.55 \times 10^{-5}$	&	$3.63 \times 10^{-5}$	&	$3.47 \times 10^{-5}$	&	$3.85 \times 10^{-5}$	&	$3.24 \times 10^{-5}$	&	19	\\
S		&	$1.62 \times 10^{-5}$	&	$1.58 \times 10^{-5}$	&	$1.55 \times 10^{-5}$	&	$1.62 \times 10^{-5}$	&	$1.31 \times 10^{-5}$	&	24	\\
Ar		&	$3.63 \times 10^{-6}$	&	$2.51 \times 10^{-6}$	&	$3.55 \times 10^{-6}$	&	$3.57 \times 10^{-6}$	&	$2.51 \times 10^{-6}$	&	45	\\
Ca		&	$2.29 \times 10^{-6}$	&	$2.23 \times 10^{-6}$	&	$2.19 \times 10^{-6}$	&	$2.33 \times 10^{-6}$	&	$2.19 \times 10^{-6}$	&	6	\\
\FM{Cr}		&	$4.68 \times 10^{-7}$	&	$4.68 \times 10^{-7}$	&	$4.47 \times 10^{-7}$	&	$5.06 \times 10^{-7}$	&	$4.37 \times 10^{-7}$	&	16	\\
\FM{Mn}		&	$2.45 \times 10^{-7}$	&	$2.45 \times 10^{-7}$	&	$3.16 \times 10^{-7}$	&	$3.56 \times 10^{-7}$	&	$2.69 \times 10^{-7}$	&	45	\\

Fe		&	$4.68 \times 10^{-5}$	&	$3.16 \times 10^{-5}$	&	$2.95 \times 10^{-5}$	&	$3.27 \times 10^{-5}$	&	$3.16 \times 10^{-5}$	&	59	\\
Ni		&	$1.78 \times 10^{-6}$	&	$1.78 \times 10^{-6}$	&	$1.66 \times 10^{-6}$	&	$1.89 \times 10^{-6}$	&	$1.66 \times 10^{-6}$	&	14	\\

\hline                                   
\end{tabular}
\par\end{centering}
\end{table}

In X-ray astrophysics, and particularly in the case of the ICM, the 
\VB{definitions} of metallicity and abundances 
\VB{presented in Eqs.~(\ref{eq:metallicity}) and~(\ref{eq:abundance}) are}
slightly adapted. 

\begin{itemize}
    \item The abundance of an element X is defined \textit{linearly},  
    with respect to H, and is noted (without brackets) as
    \begin{equation}\label{eq:abund_ICM}
        \mathrm{X} = \frac{N_\mathrm{X} / N_\mathrm{H}}{N_{\mathrm{X},\odot} / N_{\mathrm{H},\odot}}.
    \end{equation}
    \item \FM{Besides the absolute abundance of an element X (i.e. defined with respect to H), one can also consider different abundance \textit{ratios}. A common quantity to consider in this field is X/Fe, i.e. the ratio of the X abundance over that of Fe:}
    \begin{eqnarray}\label{eq:abund_ICM_ratios}
        \mathrm{X/Fe} &=& \left( \frac{N_\mathrm{X} / N_\mathrm{H}}{N_{\mathrm{X},\odot} / N_{\mathrm{H},\odot}} \right) \bigg/ \left( \frac{N_\mathrm{Fe} / N_\mathrm{H}}{N_{\mathrm{Fe},\odot} / N_{\mathrm{H},\odot}} \right)\\
        &=& \frac{N_\mathrm{X} / N_{\mathrm{X},\odot}}{N_{\mathrm{Fe}} / N_{\mathrm{Fe},\odot}}.
    \end{eqnarray}
    \FM{The interpretations of such ratios will be discussed further in this Chapter.}
    \item The metallicity $Z$ is  
    \VB{typically defined}
    with respect to the (proto-) solar metallicity $Z_\odot$. \FMbis{A common assumption when calculating metallicities in X-ray astrophysics, is that all elements have a chemical composition similar to that of our Sun (i.e. all X/Fe ratios are assumed to be 1~solar). This leaves us with only one to represent all absolute abundances at once (mostly represented by Fe in X-ray spectra; see Sect.~\ref{sec:techniques:obs}.). That variable can be then used as a one-to-one proxy of the overall metallicity of the considered object.}
\end{itemize}



\section{Stars and supernovae as sources of metals}
\label{sec:SNe}

Now that we are familiar with the concept of metallicity and abundances, one can remind fundamental basics of nucleosynthesis, i.e. the processes responsible for the transformation of light elements into heavy elements. All these elements (defined by their atomic number $\mathscr{Z}$, i.e. their number of protons) are generated from various astrophysical sources.

\FM{The two lightest elements, H ($\mathscr{Z}$=1) and He ($\mathscr{Z}$=2) were 
\VB{created} a few minutes after the Big Bang, when the Universe became cool enough to form protons and neutrons from quarks, with a fraction of them to merge directly into He nuclei. This scenario, known as the \textbf{primordial nucleosynthesis}, constitutes one of the key observational pillars in favour of the Big Bang theory (together with the expansion of the Universe and the cosmic microwave background; see also Chapter: X-ray cluster cosmology).}

\FM{Be (\FM{beryllium; }$\mathscr{Z}$=4), B (\FM{boron; }$\mathscr{Z}$=5), and to some extent Li (\FM{lithium; }$\mathscr{Z}$=3) are obtained via \textbf{nuclear spallation}, i.e. fission of heavier elements from high energy cosmic rays.}

\FM{Quite remarkably, all elements heavier than B are produced by stars and/or stellar remnants. Specifically, a star produces its internal energy by thermonuclear fusion: the extremely hot conditions in the stellar core allow pairs of H atoms to merge into He. The mass of a He atom (essentially two protons plus two neutrons) is slightly lower than the co-added mass of two H atoms (essentially two protons) and of two neutrons. The produced energy of thermonuclear fusion corresponds to this slight mass excess (transformed according to $\Delta E = \Delta m c^2$). This produced energy is necessary to compensate the star's own gravity and maintain it into hydrostatic equilibrium. Once H fusion is no more sufficient to support this equilibrium, He merges in turn into heavier elements (C, N, O, etc.) essentially via $\alpha$ process. The fate of a star will then depend on its mass, with a clear split-up between (i) low- and intermediate-mass stars ($\lesssim$8~$M_\odot$) and (ii) massive stars ($\gtrsim$10~$M_\odot$). These two separate channels will produce and release different elements, as summarised here and detailed further below:}

\begin{itemize}
    \item C ($\mathscr{Z}$=6), N ($\mathscr{Z}$=7), and to some extent Li ($\mathscr{Z}$=3) are produced by \FM{intermediate-mass stars, and more specifically by} \textbf{asymptotic giant branch (AGB) stars}.
    \item $\alpha$-elements -- \FM{products from the above mentioned $\alpha$ process;} i.e. from O ($\mathscr{Z}$=8) to Ca ($\mathscr{Z}$=20) -- \FM{are} entirely or partly \FM{produced by} \FM{massive stars and their subsequent} \textbf{core-collapse supernovae (SNcc)} explosions.
    \item Fe-peak elements -- i.e. from Cr ($\mathscr{Z}$=20) to Ni ($\mathscr{Z}$=28) -- \FM{are} largely \FM{produced} from \textbf{Type Ia supernovae (SNIa)} explosions\FM{, i.e. after low- and intermediate-mass stars terminate their life into white dwarfs}. In addition, SNIa also synthesise part of the lighter elements (typically from Si to Ca).
\end{itemize}

\FM{Last but not least,} heavier elements ($\mathscr{Z}\ge$29) are not produced by stars or stellar remnants directly. Instead, they are generally thought to originate \FM{mostly} either from \textbf{neutron star collisions} (via rapid neutron capture \FM{-- a.k.a. the r-process}) or from AGB stars (via slow neutron capture \FM{-- a.k.a. the s-process}).

\FM{Due to the specific energies and emissivities of their main transitions, o}nly elements from C to Ni are detectable in the X-ray band \FM{with current instruments. Therefore,} the sources of metals we will consider in the rest of this Chapter are AGB stars, SNcc, and SNIa. Although the following subsections further describe each of these sources, we kindly refer the reader to 
\VB{comprehensive}
reviews on stellar nucleosynthesis by \citet{karakas2014} and \citet{nomoto2013}\FM{, as well as to Chapter: Stellar evolution, supernova explosion, and nucleosynthesis}.

\subsection{Asymptotic giant branch stars}

The AGB, formally defined as a specific region on the Hertzsprung-Russell diagram, is a phase encountered by all intermediate-mass stars (1--8~$M_\odot$) at the end of their life, when their core temperature allows He to burn rapidly into C and N. During this phase, the star experiences subsequent losses of its mass via outflowing winds. These stellar winds significantly contribute to the enrichment of freshly processed C and N toward (and beyond) the local interstellar medium. The \textbf{yields} -- i.e. the mass of synthesised and ejected elements -- depend essentially on (i) the mass of the star at its birth $M_*$, and (ii) its initial metallicity $Z_\mathrm{init}$ (i.e. the amount of metals that were already present in the star at its birth)\footnote{Generally speaking, the initial metallicity of a star is a good indicator of the epoch of its formation. For instance, stars with rather high metallicities (close to our Sun: $Z_\mathrm{init} = 0.01$--0.02) formed from a more enriched gas, hence were born more recently than low-metallicity stars in, e.g., the Milky Way halo.}. Instead of one single AGB star, one might as well consider a homogeneous stellar population of common initial metallicity (i.e. born at the same time) though with diverse initial masses.
\VB{This is referred to as ``simple stellar population'' (SSP) and can be used as a very first approximation to represent a population of stars in e.g. a globular cluster, a galaxy, or a galaxy group/cluster (see also Sect.~\ref{sec:techniques:sim}).} 
\VB{In the case of an SSP,}
one can integrate the above yields over a representative distribution of their stellar masses \FM{(i.e., the initial mass function - IMF - see Sec.~\ref{sec:techniques:sim} for further details and examples)}.

\subsection{Core-collapse supernovae}

\VB{A massive star ($\gtrsim 10~M_\odot$)} 
explodes into a SNcc when it reaches the end of its ``short'' life (i.e. a few tens of Myr). 
\VB{This class comprises Types Ib, Ic, and II supernovae, which are classified differently following their spectral signatures (Type I supernovae have lost their H envelope while Type II supernovae do show H lines), but they do share the same astrophysical origin.}
Unlike low- and intermediate-mass stars, the hot core of massive stars allows heavier elements to be synthesised, and onion-like shells of burning elements progressively take shape from the surface (H burning) down to the centre (burning of heavier elements). When the core produces Fe, however, the thermonuclear fusion process becomes endothermic; i.e. it consumes energy rather than gaining it. \FMbis{The star's} gravity suddenly wins over \FMbis{its internal} pressure, the star collapses, bounces on its core (as the latter rapidly reaches the neutron degeneracy state) and violently ejects its upper layers. Freshly ejected metals are produced not only from the star before its explosion, but also (and especially) during the explosion itself. While upper layers are disseminated in the outer space, most heavy, Fe-peak elements remain locked in the core remnant, ending up in either a neutron star or a black hole \VB{(depending essentially on the intial stellar mass)}. Like for AGB stars, SNcc yields depend essentially on $M_*$ (hence, on the overall IMF) and $Z_\mathrm{init}$. A few examples of these yields are shown in the top panel of Fig.~\ref{fig:yields}. One can note (i) the particularly high yields of $\alpha$-elements compared to those of e.g. Fe and Ni, and (ii) how different models predict different yields for each individual element.

\subsection{Type Ia supernovae}

Unlike SNcc, which originate from massive stars, SNIa explosions imply low-mass stars within a binary system. More specifically, a SNIa is the result of a thermonuclear explosion of a carbon-oxygen white dwarf\footnote{A white dwarf is the compact remnant of a low-mass star ($\lesssim$8~$M_\odot$) after its final planetary nebula stage \FM{(see Chapter: White dwarfs)}.}. While in a perfect gas at fixed volume the temperature and pressure balance each other ($PV = nRT$), in white dwarfs the matter is degenerate. 
\VB{As a consequence,} if the temperature increases up to the C combustion threshold, a chain reaction will trigger the explosive burning of elements in the white dwarf, up to (and especially) its Fe-peak elements, leaving no remnant after the explosion. This way, SNIa act like thermonuclear bombs, and the energy of their explosion (i.e. their luminosity) should not depend on their intrinsic properties.  
\VB{For this reason, they are} useful standard candles to estimate cosmic distances accurately \FM{(see also Chapter: X-ray cluster cosmology)}. 

What is not clear yet, however, is the exact scenario preceding the SNIa explosion. Is the white dwarf steadily accreting material from a main sequence companion until reaching the C burning temperature (the ``single degenerate scenario'') or is this burning triggered by a violent merger between two white dwarfs (the ``double degenerate scenario'')? Whereas the former scenario is generally expected for white dwarf explosions close to the Chandrasekhar mass ($\sim$1.4~$M_\odot$; near-Chandrasekhar models), the latter is expected to occur at lower mass (sub-Chandrasekhar models). A complete review on this unsolved issue can be found in \citet{maoz2014}. Another debate stands on the physics of the explosion itself. In fact, while the general consensus is that the burning flame starts to propagate subsonically, it is not clear whether it always remains so (i.e. ``deflagration'' models) or whether it reaches a supersonic velocity before disintegrating the white dwarf (i.e. ``delayed-detonation'' models).

\begin{figure}[h]
 \centering
     \includegraphics[width=0.8\textwidth]{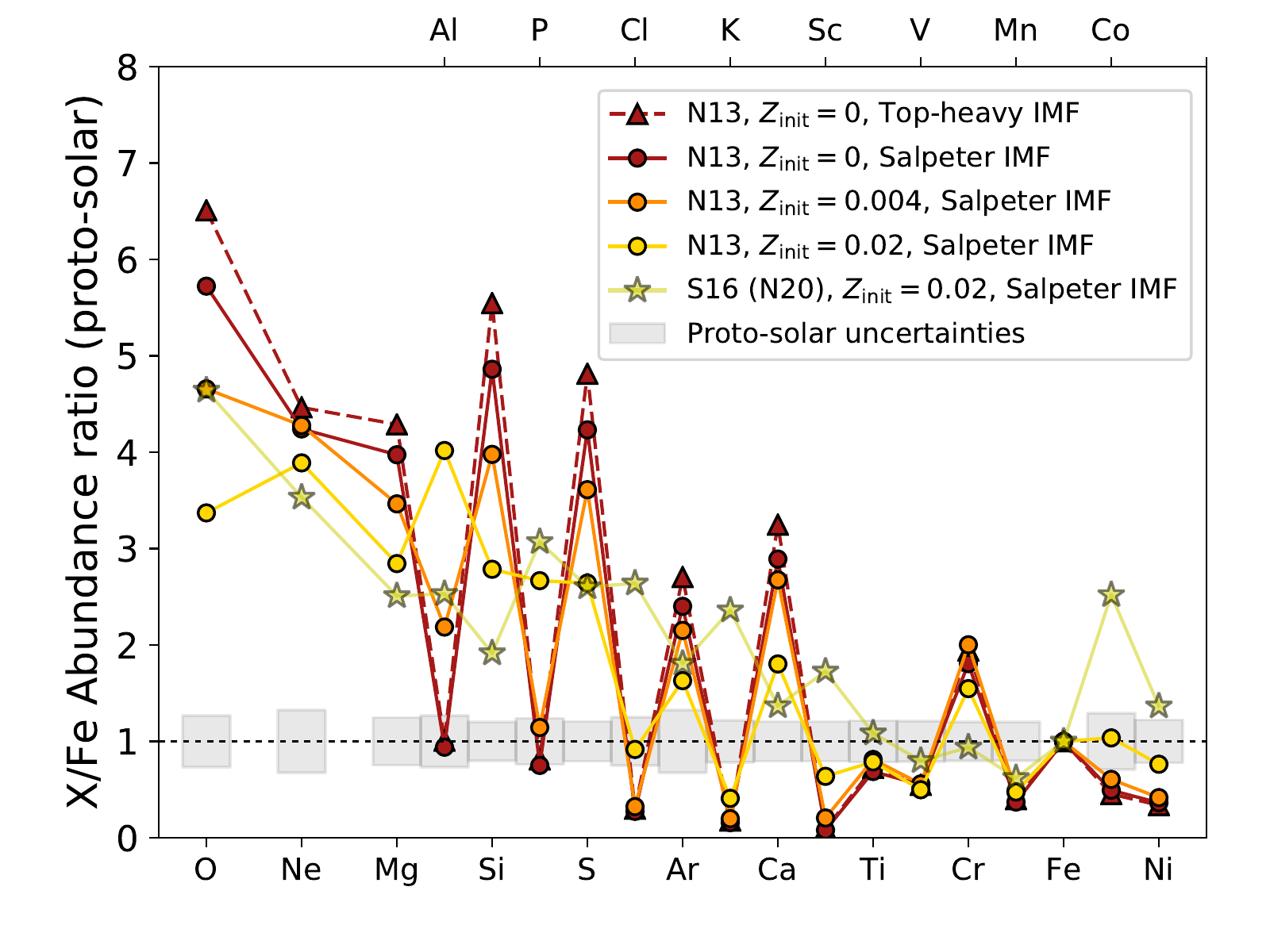} \\
     \includegraphics[width=0.8\textwidth]{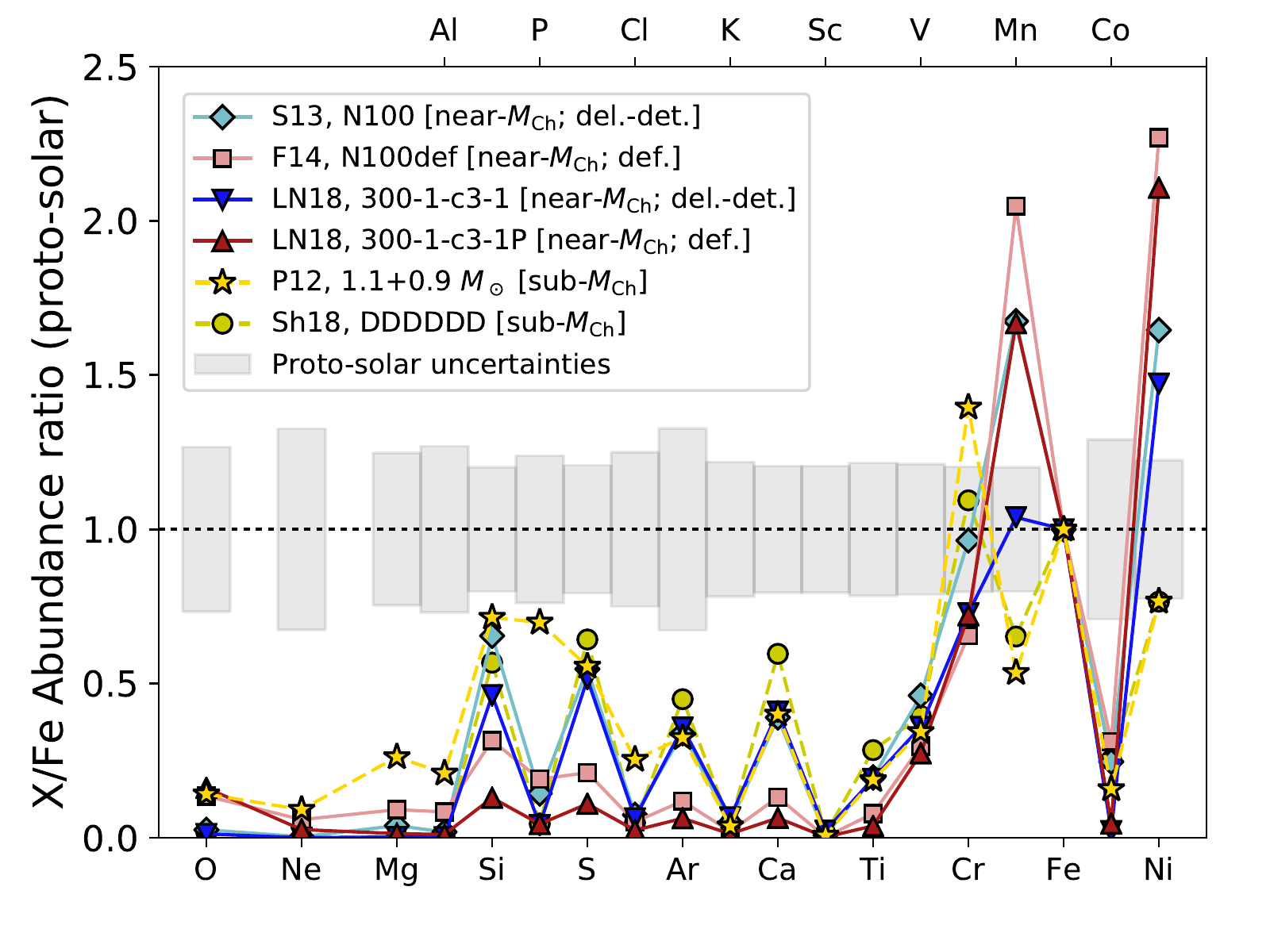}
     \caption{Yields from a set of various SNcc (\textit{top}) and SNIa (\textit{bottom}) nucleosynthesis models from the literature. Models N13 and S16 for SNcc are taken respectively from \citet{nomoto2013} and \citet{sukhbold2016}, and have been integrated by mass over a 
     \VB{Salpeter~\cite[][]{salpeter1955} or a top-heavy~\cite[e.g.][]{arimoto1987} IMF}. Models S13, F14, LN18, P12, and S18 for SNIa are taken respectively from \citet{seitenzahl2013}, \citet{fink2014}, \citet{leung2018}, \citet{pakmor2012}, and \citet{shen2018}. Reprinted with permission from \citet{mernier2018c}.}
     \label{fig:yields}
 \end{figure}

Interestingly, each of these models predicts different yields. This is shown in the bottom panel of Fig.~\ref{fig:yields} where 
differences between deflagration vs. delayed-detonation models (as well as near- vs. sub-Chandrasekhar models) clearly  
\VB{arise, despite a convergent view that SNIa produce substantial amounts of Si, S, Ar, Ca, Cr, Mn, Fe, and Ni.}
Note that, although deflagration models on the figure seem to produce larger amounts of Ni (with respect to Fe), updated calculations suggest that this is not necessarily the case \citep[][see also Sect.~\ref{sec:stellar_physics}]{nomoto2018,simionescu2019}.

\section{Measuring/simulating the ICM chemical properties: techniques and current limitations}
\label{sec:techniques}

\subsection{Deriving abundances from X-ray spectroscopy}
\label{sec:techniques:obs}

 
\VB{The} presence of metals in the ICM is revealed  
by prominent emission lines  
\VB{over} the spectral continuum \VB{of the ICM emission}. These lines correspond to specific (K-shell and L-shell\footnote{A K-shell transition occurs from a given ion when an electron falls to the $n=1$ (1s) orbital from a higher energy level, thereby emitting an X-ray photon of that specific energy. On the other hand, a L-shell transition occurs when an electron falls to the $n=2$ (2s or 2p) orbital from a higher energy level.}) atomic transitions of electrons onto heavily ionised elements. 

In a collisional plasma such as the ICM, the continuum  
\VB{of the spectrum predominantly originates from the
\textit{bremsstrahlung} emission that occurs}  
when a free electron passes close to a free ion, interacts with its electrostatic field, loses kinetic energy by deviating from its initial trajectory, and emits an X-ray photon to conserve the total energy. This is thus a ``free-free'' interaction, and consequently bremsstrahlung photons can be emitted over a continuous range of energies \citep{sarazin1986,boehringer2010}. The (X-ray) emissivity $\epsilon_\mathrm{cont}$ of this process depends on (i) the gas electron density $n_e$\footnote{For a fully-ionised plasma like the ICM, the electron density is typically supposed to be $n_e \simeq 1.18\, n_i$, \FMbis{where $n_i$ is the proton (or ion) density of the plasma}. The total density is thus $n_\mathrm{tot} = n_e + n_i \simeq 1.85\, n_e$.}, (ii) the gas electron temperature $kT$\footnote{In X-ray astrophysics, the temperature $T$ is often multiplied with the Boltzman constant $k$, and thus expressed in units of keV (kilo-electronvolt).} (in other words, on the average incident kinetic energy of the free electrons), and (iii) the emitted photon energy $E$. More specifically, it is proportional to
\begin{equation}\label{eq:bremsstrahlung}
    \epsilon_\mathrm{cont}(E) \propto \bar{g} n_e^2 (kT)^{-1/2} e^{-E/kT},
\end{equation}
where $\bar{g}$ is the Gaunt factor and depends weakly on the photon energy. From Eq.~(\ref{eq:bremsstrahlung}), one can easily see that the spectral shape depends on the temperature: photons of higher energies are enhanced with hotter gas. \FM{ Generally speaking, this continuum emissivity scaled as $\propto n_e^2$ is often directly associated with the so-called gas emission measure (for a concrete application, see Sect.~\ref{sec:techniques:sim}). At a weaker level, since the $\propto n_e^2$ dependency originates from the product of the electron \textit{and} proton number densities, the continuum emissivity slightly depends on the abundance of ion species as well. As for the temperature dependency, a concrete example is discussed further below.}

In addition to the \VB{continuous} free-free emission, X-ray photons are  
also emitted at very specific energies, corresponding to discrete transitions of bound electrons onto specific ions. These transitions often occur after a free electron collides with an atom, repelling a bound electron to an orbital of higher energy. In reaction, this bound electron will cascade down again to an orbital of lower energy by emitting an X-ray photon corresponding to the energy difference $\Delta E$ of the transition \citep{sarazin1986,boehringer2010}. In practice, modelling these atomic transitions requires complex calculations which are incorporated in publicly available spectral fitting packages (e.g. XSPEC\footnote{\url{https://heasarc.gsfc.nasa.gov/xanadu/xspec/}}~\cite{xspec}, SPEX\footnote{\url{https://www.sron.nl/astrophysics-spex/}}~\cite{spex}). One can show, however, that the integrated line emissivity $\epsilon_\mathrm{line}^\mathrm{int}$ for an ion $i$ of an element X is proportional to
\begin{eqnarray}\label{eq:line}
    \epsilon_\mathrm{line}^\mathrm{int} &=& \int \epsilon_\mathrm{line}(E) dE \\
    &\propto& n_e^2 \left( \frac{N_{\mathrm{X},i}}{N_\mathrm{X}}\right) \left( \frac{N_\mathrm{X}}{N_\mathrm{H}}\right) (kT)^{-1/2} E \Omega(T) e^{-\Delta E/kT},
\end{eqnarray}
where $\Omega(T)$ is the collision strength (which varies weakly with temperature), $(N_{\mathrm{X},i}/N_\mathrm{X})$ is the ion number fraction -- i.e. the fraction of the considered ion over the total number of atoms of that species, and $(N_\mathrm{X}/N_\mathrm{H})$ directly relates to the abundance of element X (Eq. \ref{eq:abund_ICM}). In the case of a tenuous plasma like the ICM, the ion fraction depends only on the temperature and is independent of the density. Thus, at fixed temperature, both the continuum and (integrated) line emissivities display the same square dependency on the electron density. It  
\VB{follows} that the \textit{equivalent width} of a given emission line (i.e. the ratio of its emissivity $\epsilon_\mathrm{line}^\mathrm{int}$ over that of the continuum $\epsilon_\mathrm{cont}$ at the line energy) is directly proportional to the abundance of its element: the more a line \FM{spikes} above the continuum, the more abundant  
\VB{the} associated element~is.

\begin{figure}[!]
 \centering
     \includegraphics[width=0.85\textwidth, trim={0.cm 1.8cm 0.cm 0.3cm},clip]{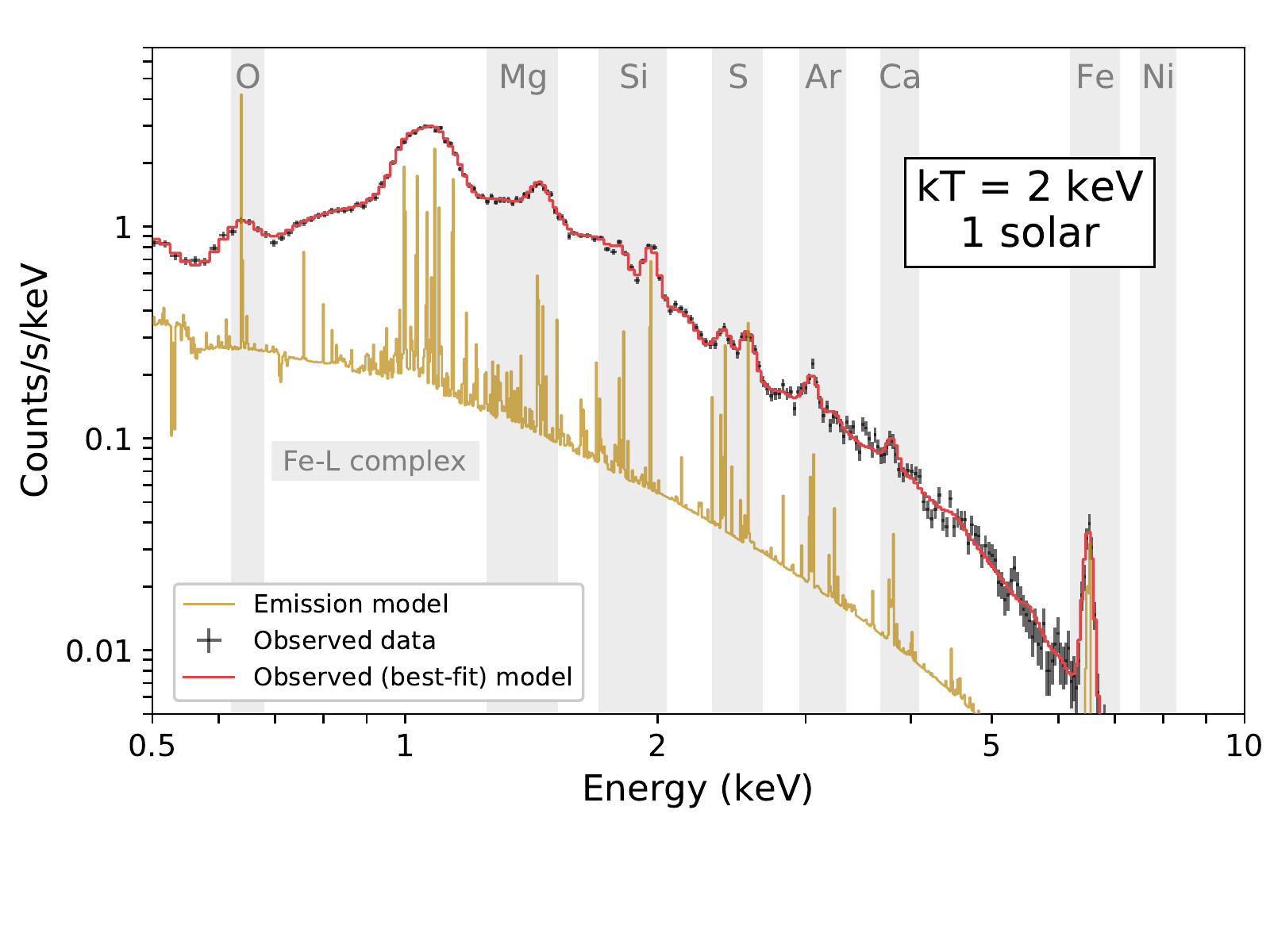} \\
     \includegraphics[width=0.85\textwidth, trim={0.cm 1.8cm 0.cm 0.3cm},clip]{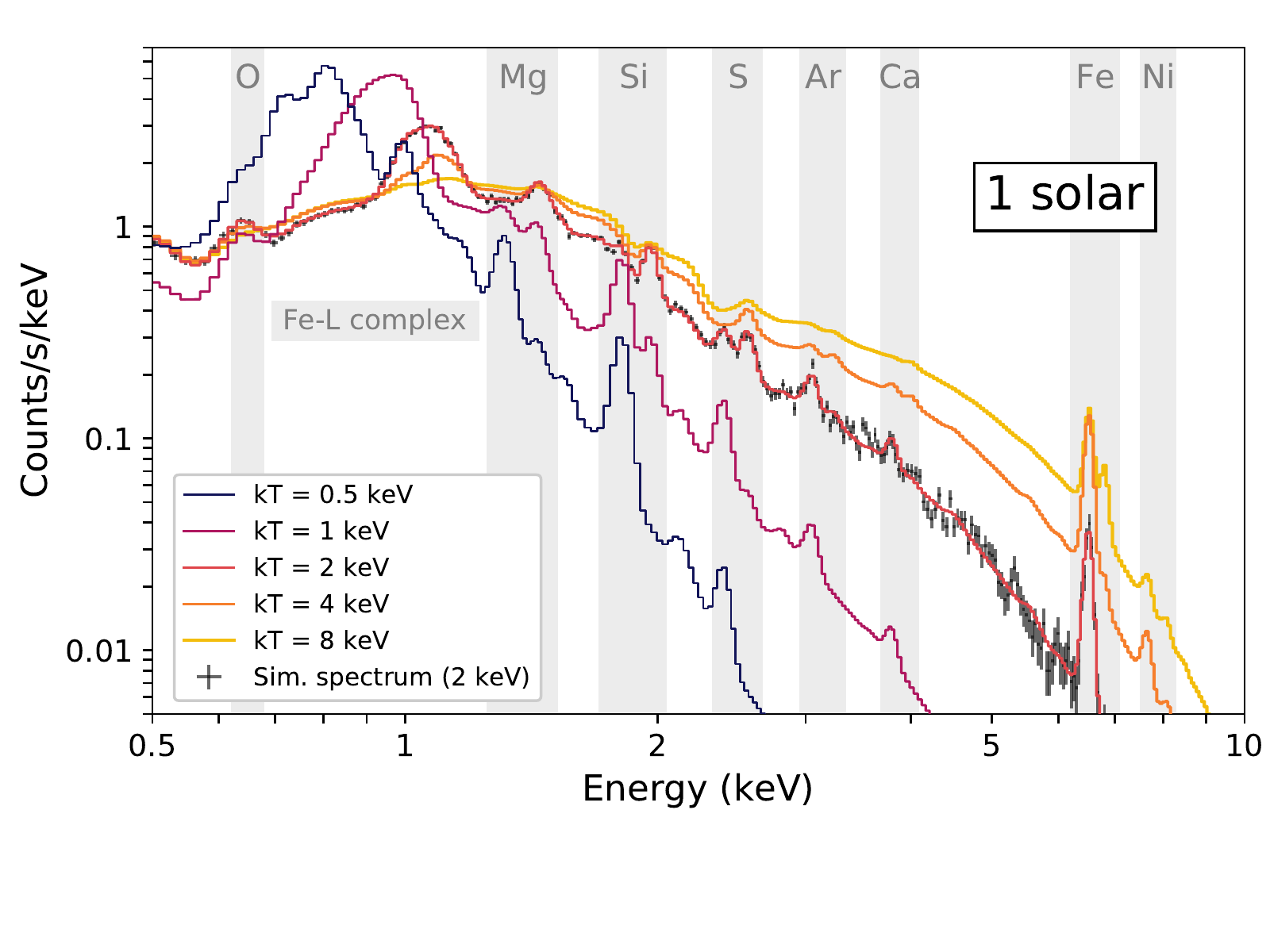} \\
     \includegraphics[width=0.85\textwidth, trim={0.cm 1.8cm 0.cm 0.3cm},clip]{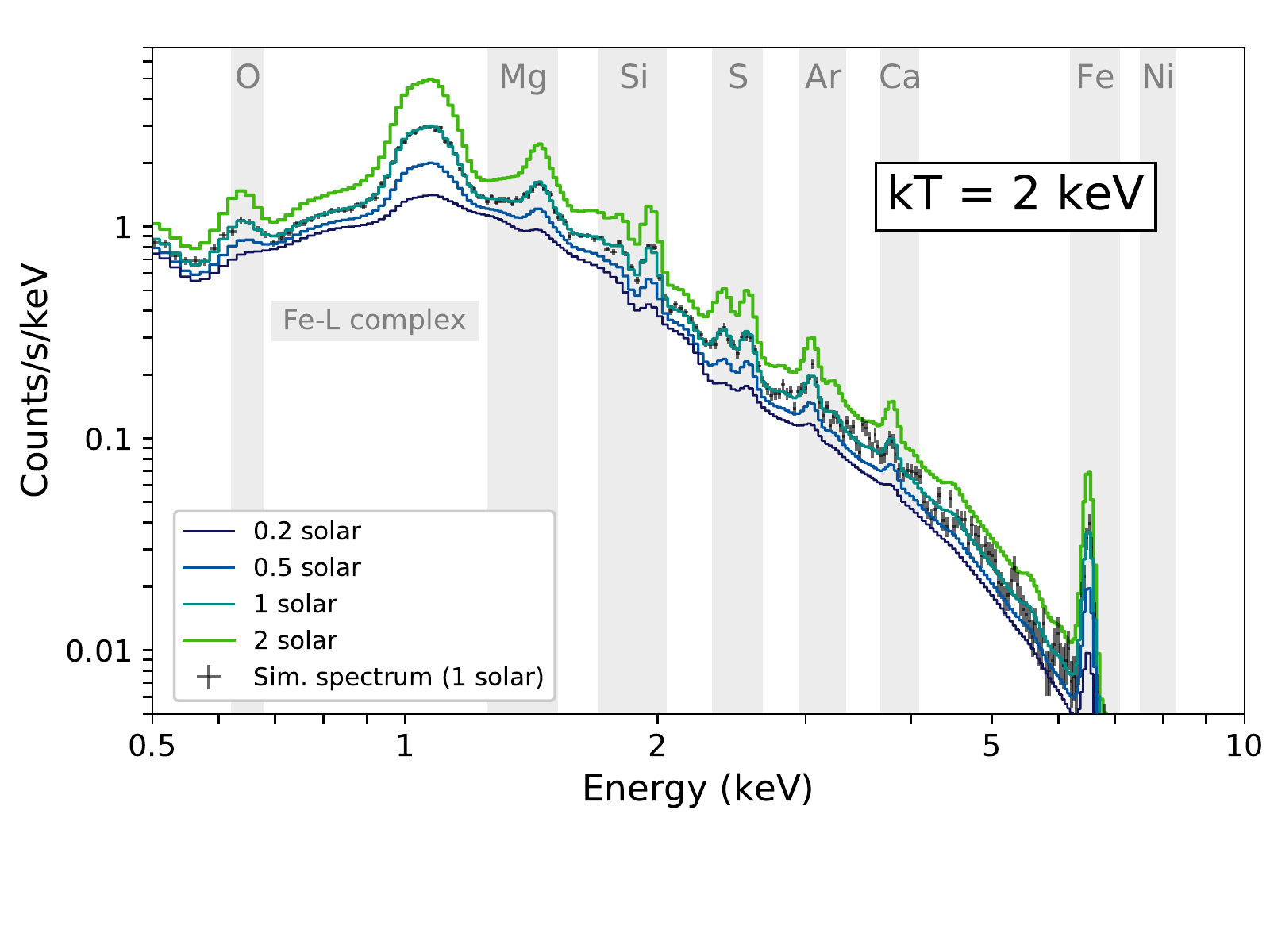}
     \caption{\textit{Top:} Theoretical spectral emission model for a $kT = 2$~keV gas assumed with proto-solar abundances (in units of L09). Compared to its associated mock \textit{XMM-Newton}/MOS observed spectrum (simulated here with 50~ks), the model has been shifted low for clarity. \textit{Middle:} Set of MOS mock spectra with various gas temperatures (from 0.5 to 8~keV) at fixed solar abundances. \textit{Bottom:} Set of MOS mock spectra with various abundances (from 0.2 to 2~solar) at fixed $kT = 2$~keV. For easier comparison, (simulated) spectral data in all panels are drawn for the same model ($kT = 2$~keV, solar abundances).}
     \label{fig:spectra}
 \end{figure}
 
\FM{Let us now illustrate these two spectral features (i.e. continuum and line emission) with a concrete example.} The top panel of Fig.~\ref{fig:spectra} shows a theoretical X-ray spectrum of a typical poor cluster (or rich group) with $kT = 2$~keV and proto-solar abundances (in units of L09), calculated using a \texttt{cie} model from the spectral code SPEX v3.06 (brown-yellow, shifted low for clarity). This cluster has been assumed to be at a distance of 100~Mpc (i.e. $z \simeq 0.02$) and to be absorbed by a Galactic hydrogen column density of $n_H = 10^{21}$ particles/cm$^2$. In reality, however, spectra are never observed as such: X-ray optics and instruments have a limited spectral resolution and their effective area (i.e. sensitivity) varies with energy. Therefore, such an idealised spectrum needs to be convolved with the response of the instrument, in order to generate a model \FM{(red line)} that can be directly compared to the observed spectral data (black data points; here simulated for 50~ks of exposure with the \textit{XMM-Newton}/MOS instrument). Instruments onboard X-ray observatories (especially at the current CCD-like resolution \VB{of the order of $\sim 120\,$eV}) thus unavoidably transform lines into less resolved ``bumps''. Nevertheless, these bumps can still be reliably converted into abundances, as most transitions are well separated in energy for each element (gray vertical strips in the figure). One notable exception is the so-called ``Fe-L complex'' between 0.7--1.3~keV, consisting of a large number of L-shell Fe (and some Ni) transitions, as well as K-shell Ne transitions. The loss of spectral information is more important in that complex; therefore Fe, Ne, and Ni abundances estimated from there should be treated with extreme caution. It also becomes clear that, when using a model that measures the whole metallicity (i.e. when the abundance of all metals are tied together \FMbis{-- see Sect.~\ref{sec:formalism}), like} the \texttt{apec} model in XSPEC,  
\VB{its value}
will essentially trace the Fe abundance. This is because the high statistics of the Fe-L complex prompts the fit to reproduce this bump in priority compared to the K-shell transition of other elements. Moreover, \FMbis{with the notable exceptions of C, N, and O (which are not always easy to constrain in the ICM),} at similar abundances the mass of Fe usually dominates that of other elements.

Starting from the (mock) observed spectrum and model described above, we now explore how its spectral shape changes with temperature. This is shown in the middle panel of Fig.~\ref{fig:spectra}, where we simulate the same ICM at four other gas temperatures. For clarity, each model is simulated at constant abundances (1~proto-solar in units of L09) and 
rescaled for their flux to be unchanged within the 0.3--2~keV energy band. As explained above, one can clearly see that the \FMbis{exponential cut-off of the bremsstrahlung emission occurs at higher energies with increasing temperature, translating into a flatten spectral slope for hotter clusters}. Another notable feature is that, at fixed abundances, the equivalent width of lines at lower energies tends to weaken with increasing temperature. This is because at higher temperatures light atoms, i.e. with smaller binding energies, are more easily fully ionised, with less bound electrons available for atomic transitions. On the contrary, transitions requiring higher collision energies (in particular K-shell transitions of heavy ions) take over in hotter plasmas. A striking case is that of Fe. In galaxy groups (i.e. $kT \lesssim 1$~keV), the Fe abundance is essentially determined through the Fe-L complex. At the higher temperature regime of galaxy clusters, however, more Fe-K transitions are available and thus Fe-K lines around $\sim$6.7~keV become much easier to detect. Thus, in clusters, the Fe abundance is essentially measured via this Fe-K bump.

The last example consists in the opposite exercise, i.e. keeping a fixed temperature ($kT = 2$~keV) while varying the metallicity. This is shown in the lower panel of Fig.~\ref{fig:spectra}. As explained above, one can easily see how higher abundances translate into larger equivalent widths of their corresponding line emission. It then becomes clear that, when the exposure of a system is deep enough, its statistics allow to constrain key parameters for corresponding up-to-date spectral models (via simple fitting), thus to derive not only its ICM temperature but also its chemical abundances.

\subsection{Current observing limitations}

As we have seen, the limited spectral resolution of CCD-like spectrometers onboard currently flying X-ray observatories (for instance \textit{XMM-Newton}/MOS in our exercise above) implies that lines are often blended. This is especially true for the Fe-L complex, whose (unresolved) line emission may largely dominate the whole spectrum in the groups regime. Consequently,  
\VB{a degeneracy arises}
between the metallicity and the emission measure (directly related to the gas density) of cool systems -- typically at $kT \lesssim 0.7$~keV -- observed with shallow exposures.

Another systematic effect is the so-called ``Fe-bias'' \citep{buote1998,mernier2018c,gastaldello2021}: it occurs when an actual multi-temperature ICM is fitted with a single-temperature plasma model only. In fact, projection effects as well as intrinsic thermodynamics often imply a complex temperature structure of the ICM. In that case, the addition of lower and higher temperature components will boost Fe-L transitions across a broad energy window, resulting in a ``flatter'' \FM{bump forming} the Fe-L complex (see e.g. differences within $ 0.5~\mathrm{keV}\le kT \le 2~\mathrm{keV}$ in Fig.~\ref{fig:spectra}, middle panel).   
\VB{As the determination of the continuum is very problematic, modelled}
with one single-temperature component only,
the fit will incorrectly interpret this flatter \FM{bump} as a gas having a low Fe abundance (see e.g. the 0.2--0.5 solar cases in Fig.~\ref{fig:spectra}, bottom panel). A bias with similar origin, called the ``inverse Fe-bias'', has also been found in slightly hotter clusters ($kT \sim 2-3$~keV), in which the Fe abundance is derived from \textit{both} the Fe-L complex and the Fe-K lines. If the gas is multi-temperature while modelled with one temperature only, its Fe abundance becomes then overestimated compared to reality \citep{rasia2008,mernier2018c,gastaldello2021}.

Other limitations may also \FMbis{arise} from systematic uncertainties that are either from astrophysical origin (contamination from e.g. the cosmic X-ray background, emission from neighbouring AGN or other X-ray sources, uncertainties on the foreground Galactic absorption, etc.) or from instrumental origin (calibration issues, uncertainties on the instrumental background, etc.).

Clearly, the key direction to minimise most of these limitations is to 
\VB{considerably improve} the spectral resolution of our X-ray instruments. This is already the case, for instance, with the RGS instrument onboard \textit{XMM-Newton}. On paper, this dispersive instrument (made of gratings decomposing the X-ray light along a dedicated detector\FM{; see Chapter: Gratings for X-ray astronomy}) has the potential to improve its resolution by up to a factor $\sim$40 compared to CCD-like spectroscopy. In practice, however, the spatial extent of clusters and groups results in a blurring of all features in RGS spectra, hence somewhat decreasing its resolving power. Moreover, RGS cannot perform spatial spectroscopy, what somewhat limits its scientific potential for this field. A couple of interesting and valuable abundance studies made with RGS, however, are available in the literature \citep[e.g.][]{werner2006,deplaa2017,mao2019}. Future missions focusing on non-dispersive high-resolution spectroscopy are detailed further in Sect.~\ref{sec:future}.

\subsection{Simulations}
\label{sec:techniques:sim}

The chemical enrichment of the ICM in clusters of galaxies can also be explored from a theoretical point of view. 
Typically, this is done via numerical models of cosmic structures, such as groups and clusters, either with semi-analytic models (SAM; e.g.\ early works by~\citet{delucia2004,nagashima2005} and recent investigations by~\citet{yates2017}) or through cosmological hydrodynamical simulations~\cite{borgani2008,biffi2018b}. 
In addition to the gravitational force, a vast range of non-linear physical processes drives the formation and evolution of clusters in the Universe, and a numerical approach is in fact required to model all of them. 
Indeed, most of the baryonic processes, from star formation to thermal and chemical feedback, are strongly interconnected and influence the observable properties of clusters, such as their metal enrichment level.

Both theoretical approaches, namely SAM and hydrodynamical simulations, have advantages and disadvantages. 
SAM start from ``dark matter only'' simulations and populate dark matter haloes with galaxies, whose evolution and properties are described with analytic formulae and can be 
\VB{calibrated} to reproduce observational evidence. These models are therefore computationally cheap, but they rely on a number of assumptions and free parameters.
On the other hand, numerical simulations are computationally much more expensive to perform because they follow in detail the gravitational and hydrodynamical forces, and describe the evolution of the baryonic component by starting from the equations of hydrodynamics.  
\VB{In addition,}
given the wide dynamical range involved in the description of galaxy clusters (from several Mpc for gravitational processes to sub-parsec scales for star formation), numerical simulations must resort to so-called \textit{sub-resolution} \FM{({a.k.a. sub-grid})} models to describe most of the baryonic processes.

In this Chapter we will mainly focus on results from hydrodynamical simulations, in which the enrichment of the ICM in clusters/groups is treated by embedding stellar evolution models for metal production and release. \FM{T}he evolution of the stellar component and chemical enrichment of the surrounding medium \FM{rely on} \FM{complex physical} processes happening on \FM{\VB{very different} spatial and \VB{time}} scales \VB{with respect to large-scale processes like cluster formation and assembly. Thus they cannot}  
be directly resolved in a simulation of a large cosmological volume -- necessary to follow the formation of \VB{massive cosmic structures}. 
An interesting approach is to simulate galaxy clusters with \textit{zoom-in} techniques, in which higher resolution is concentrated on the high-density region where the cluster is located while the large-scale background is described with coarser resolution. Even in this case, it is however not possible to resolve the formation and evolution of a single star from the cooling of a gas cloud, and a sub-resolution description is thus required.
Typically, simulations of galaxy clusters approximate the stellar component with resolution elements that have masses of order \VB{$10^4$--$10^7\,M_\odot$.}
Therefore, each stellar element rather represents a population of stars, all characterised by the same age and initial metallicity (i.e. a simple stellar population, SSP -- see also Sect.~\ref{sec:SNe}), and by an assumed IMF (see below).

The most detailed treatments of chemical enrichment included into numerical simulations take into account stellar evolution models to compute the expected amount of metals released by every stellar element in the simulation through the main production channels, namely SNIa, SNcc and AGB stars.
These models allow to follow the chemical enrichment of the cosmic gas through the evolution of the Universe in simulations of large cosmic volumes, isolated galaxy clusters and groups, and galaxies.
The chemical enrichment modelling requires some fundamental assumptions to describe the evolution of the SSP that each stellar element represents. 
In particular,
\VB{these models are based on the fundamental pillars already introduced in this Chapter -- namely the IMF, stellar lifetime function and metal yields. We \FMbis{review} them in the following, in light of their numerical modelisation within chemical models.}
\begin{itemize}
    
    \item \textbf{the \FM{initial mass function (IMF)}} ($\phi(m)={\rm d}N/{\rm d} m$), which is the birth mass distribution of stars and is defined as 
    \FM{the number of stars at a given mass}
    (see also Sect.~\ref{sec:SNe}). This essentially establishes the relative \FMbis{fractions} of (long-living) low-mass stars and \FMbis{of} (short-living) high-mass stars. In particular, the shape of the IMF predicts also the fractions of subsequent SNIa and SNcc, which is in turn connected to the relative ratio between $\alpha$-elements and Fe-peak elements (Sect.~\ref{sec:SNe}).
    A very common form of the IMF is for instance the one proposed by~\citet{salpeter1955}, which is characterised by a single power-law shape $\phi(m)\propto m^{-2.35}$.
    Other functional forms of the IMF can predict different amounts of low- or high-mass stars, such as the (flatter) top-heavy IMF proposed by~\citet{arimoto1987} $\phi(m)\propto m^{-1.95}$.
    The IMF can also be characterised by different slopes and shapes in different mass regimes, such as those introduced by~\citet{kroupa2001} or~\citet{chabrier2003}.
    Differences between a few examples of IMFs widely adopted in the literature are shown in Fig.~\ref{fig:IMFs}. In particular one can note how the top-heavy IMF by~\citet{arimoto1987} predicts a larger fraction of massive stars compared to the function by~\citet{salpeter1955}, differently from the one by~\cite{kroupa1993}. At low masses, the Salpeter IMF generally over-predicts the fraction of stars if compared to the other IMFs shown here.
    
    \begin{figure}
        \centering
        \includegraphics[width=0.75\textwidth]{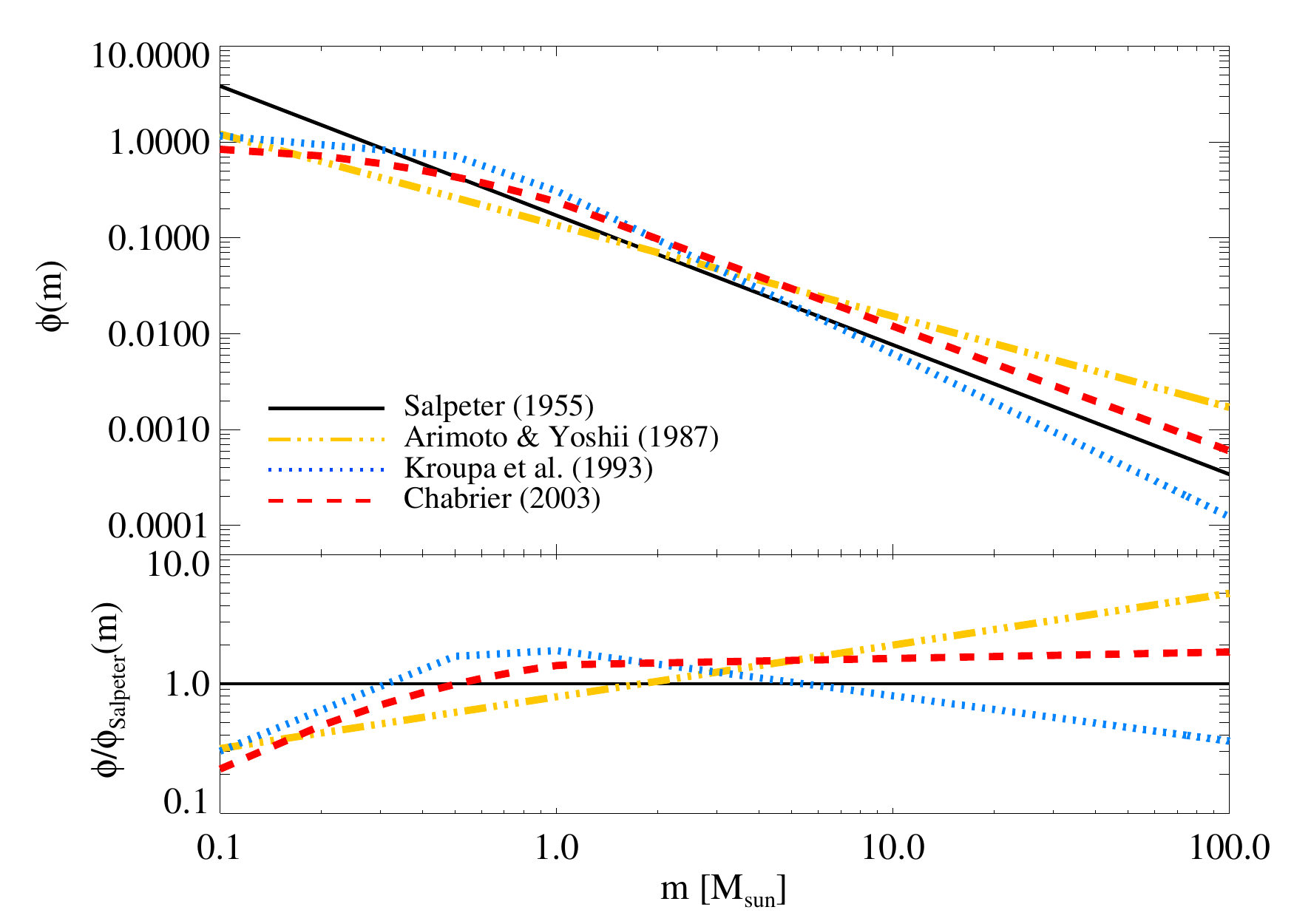}
        \caption{Comparison between four of the most common IMFs, as in the legend, as function of stellar mass. In the bottom panel we report the ratio of each IMF to the reference IMF by Salpeter~\cite[][]{salpeter1955}.  
        Reprinted with permission from~\protect\citet{biffi2018b}.}
        \label{fig:IMFs}
    \end{figure}
    
    To date there is however no general consensus on the shape of the IMF, nor on whether it is universal or dependent on time and/or environment~\cite[][]{bastian2010,kroupa2013,hopkins2018}. \FM{\VB{These fundamental questions, however, can be also addressed} 
    by measuring accurately abundances in the ICM, as the chemical signature of early stellar populations (see Sect.~\ref{sec:stellar_physics} for further detail).}
    
    \item \textbf{the lifetime function}, 
    which parametrises the typical lifetimes of stars with different masses. This also has an impact on the absolute and relative abundance of chemical elements released by different sources, because it determines the \textit{time delay} after which a star of a given mass will enter the final stages of its life ejecting metals during the the AGB phase or via SN explosion.
    
    In the literature, various choices of the mass-dependence of the lifetime function have been proposed (see~\cite{romano2005_I}, for an overview), as well as models in which the lifetime dependence on the stellar metallicity is also taken into account~\cite[e.g.][]{portinari1998}.

    \item \textbf{the stellar yields}, 
    already discussed above (Sect.~\ref{sec:SNe}), which provide the ejected amount of the different metal species by different stellar sources, depending on their initial mass and metallicity~\citep{nomoto2013,karakas2014}.
\end{itemize}

The fundamental set of integral equations describing the evolution of the density for each chemical element essentially computes the rates at which stars of different masses (and initial metallicity) in each SSP explode as SNcc or SNIa or undergo the AGB phase -- depending on their lifetimes, and the corresponding amount of ejected metal yields~\cite[][]{matteucci2003}. In addition to the contributions from each production channel (i.e. SNIa, SNcc and AGB stars), it is also important to consider the depletion of metal-rich gas into new episodes of star formation, at any given time. An example for the implementation of these equations in a cosmological hydrodynamical code is provided by the model by~\citet{tornatore2004,tornatore2007} (see also~\cite{biffi2018b}, for a recent review).

\FMbis{The} typical metallicity or chemical abundance for a specific region of the cluster can be then computed by averaging \FMbis{metals} over the gas \FMbis{elements} \VBbis{comprised in the region}. The average value can be weighted by different properties \VBbis{of the gas elements}, such as gas \textit{mass} or some quantity proportional to the expected \textit{emission} of the gas, such as the gas emission measure ($\propto n_e^2$) or, even better, the X-ray emissivity ($\propto n_e^2 \Lambda(T,Z)$, where $\Lambda$ is the cooling function and depends on gas temperature and metallicity \VB{-- \FM{thus taking into account the continuum \textit{and} line emissivities;}}). The latter is particularly useful for comparisons between numerical simulations and X-ray observations, in which the metallicity is retrieved from emission spectra of the ICM
(see Sect.~\ref{sec:techniques:obs}).

\subsection{Numerical uncertainties and limitations}

The impact of the assumptions at the base of chemical models can be significant because metals affect not only the resulting chemical properties of the ICM but also the gas cooling~\cite{sutherland1993,maio2007,wiersma2009} and star formation.
Uncertainties related to the adopted stellar yields combined to changes in the assumed IMF, for instance, can alter the resulting chemical enrichment or abundance ratios by a factor of a few by redshift $z\sim 0$~\cite[][]{vogelsberger2018}. 
Changes in the relative fraction of low- and high-mass stars predicted by different IMF not only directly affect the relative abundances of different chemical species, but also impact the stellar feedback in the form of winds driven by SNcc and massive stars. In turn, this affects the thermal properties of the ICM and the star formation history.
Given such complex interplay between the chemical enrichment and the other baryonic processes, any uncertainty on the chemical model assumptions can indeed impact the global properties of the simulated cosmic structures in a non-trivial way.
This constitutes a limitation and has to be taken into account as cluster simulations should aim at reproducing \textit{simultaneously} chemical and thermo-dynamical properties of observed clusters, and possibly the galaxy population.

From the numerical point of view, resolution and details of the hydrodynamical code used to perform the simulations can additionally affect the resulting chemical patterns.
Several studies have shown that the chemical enrichment level of the ICM generally tends to increase with increasing numerical resolution, and the enrichment history will likely change as well. Higher resolution, especially at high redshifts, will in fact enhance star formation, due to the larger number of resolved haloes that undergo collapse and gas cooling. Increasing the resolution also requires the re-calibration of the physical models included in the simulations to maintain the agreement with a few reference \VB{observables, such as the stellar mass function, the relation between super-massive BH mass and stellar mass of the host galaxy or the baryon fraction}. For this reason, the convergence on the resulting chemical properties cannot be ensured.

The hydrodynamical scheme itself can further impact the distribution of metals from stars into the surrounding gas, depending on the implementation of mixing and diffusion processes. This aspect affects different classes of simulation codes in different ways, depending on the characteristics of the approach adopted. For instance, the so-called Eulerian codes, which solve the hydrodynamics equations on a grid representing the fluid,
are characterised by an implicit treatment of numerical diffusion that can in some cases lead to an over-mixing.
In contrast, codes using a Lagrangian approach and discretising the description of the fluid in terms of particles, necessitate an explicit implementation of a diffusion model to distribute and mix the metals into the gas. A few studies have attempted to implement these diffusion processes within chemical models in particle-based codes~\cite[][]{greif2009,williamson2016}, though typically limited to the small scale over which the hydrodynamical quantities are computed. 
The transport of the metals on larger scales, however, can be relevant as well, and can also be affected by other physical processes (see also following Section~\ref{sec:when}).


\section{How and when did the ICM become chemically enriched?}\label{sec:when}

\VB{Due to the discussed difficulties related to spectral line measurements, metals are}
detected in the ICM of bright, nearby systems -- simply because these systems emit a higher number of photons (hence providing better statistics to constrain their abundances).  
\VB{The metals of this diffuse medium are those that}
have left galaxies and progressively accumulated into the gravitational well of clusters and groups over cosmic time. In the case of clusters,  
\VB{the} gravitational well is deep enough to retain all metals that have fallen into it\footnote{This concept is often reported in the literature as ``closed-box'' systems. Note that, unlike clusters, galaxy groups have a shallower gravitational well and should \textit{not} be considered as closed-box \citep[for a review specific to metals in galaxy groups, see][]{gastaldello2021}.}. In other words, abundances measured in the ICM often constitute a remarkable fossil record of chemical enrichment accumulated over time at  
\VB{large} scales of the Universe. 

From this chemical footprint, two questions emerge. First: \textit{at which cosmic epoch the bulk of metals left galaxies to enrich the ICM?} Second: \textit{which physical mechanisms are responsible for the diffusion of metals from stars to Mpc scales?} Although these questions  
\VB{tackle different issues}
at first glance, we will see that they can be answered simultaneously via observations and simulations.

\subsection{Spatial uniformity of the metal distribution}\label{sec:when:profiles}

An excellent way to investigate the history of the ICM enrichment is to explore the spatial distribution of chemical elements across the cluster extent, from the centre out to the periphery.
Early observations in the X-ray band focused mostly on the bright core of relaxed, cool-core clusters\footnote{For an extensive description of cool-core and non-cool-core clusters, we refer the reader to \FM{Chapter: Thermodynamical profiles of clusters and groups, and their evolution and Chapter: Scaling relations of clusters and groups, and their evolution}.}, extending at most to half of the virial radius, due to the rapidly decreasing X-ray surface brightness of the ICM at larger distances from the centre.
Results indicate a gradient in the Fe abundance, which decreases from the innermost regions outwards. In particular, the central Fe peak has been observed in several nearby bright cool-core clusters since the early $'90$s. Differently, non-cool-core clusters typically exhibit flatter radial profiles over the entire radial range. 
Despite their lower brightness, cluster outskirts are nonetheless very interesting regions to explore: not only they cover most of the system volume, but they also retain direct information on the mass accretion and the connection to the large scale structure~\cite[][\FM{see also Chapter: Cluster outskirts and their connection to the cosmic web}]{walker2019}. In fact, looking further away from the cluster core somehow corresponds to looking back in the ICM chemical history, as the gas in the outskirts has freshly accreted and is less well virialised and stratified than in the innermost regions. 
In the recent years, a number of X-ray observations of galaxy clusters, mostly performed with the \textit{Suzaku} satellite (thanks to its lower instrumental background)~\citep[e.g.][]{werner2013,urban2017}, allowed to map the metallicity of the ICM outside $R_{500}$, reaching in some cases the virial boundary (or $R_{200}$)\footnote{In galaxy clusters, $R_{500}$ ($R_{200}$) is defined as the radius comprising an average density which is $500$ ($200$) times the critical density of the Universe.}. 
\begin{figure}[t]
    \centering
    \includegraphics[width=1.0\textwidth, trim={1.5cm 0cm 2.3cm 1cm},clip]{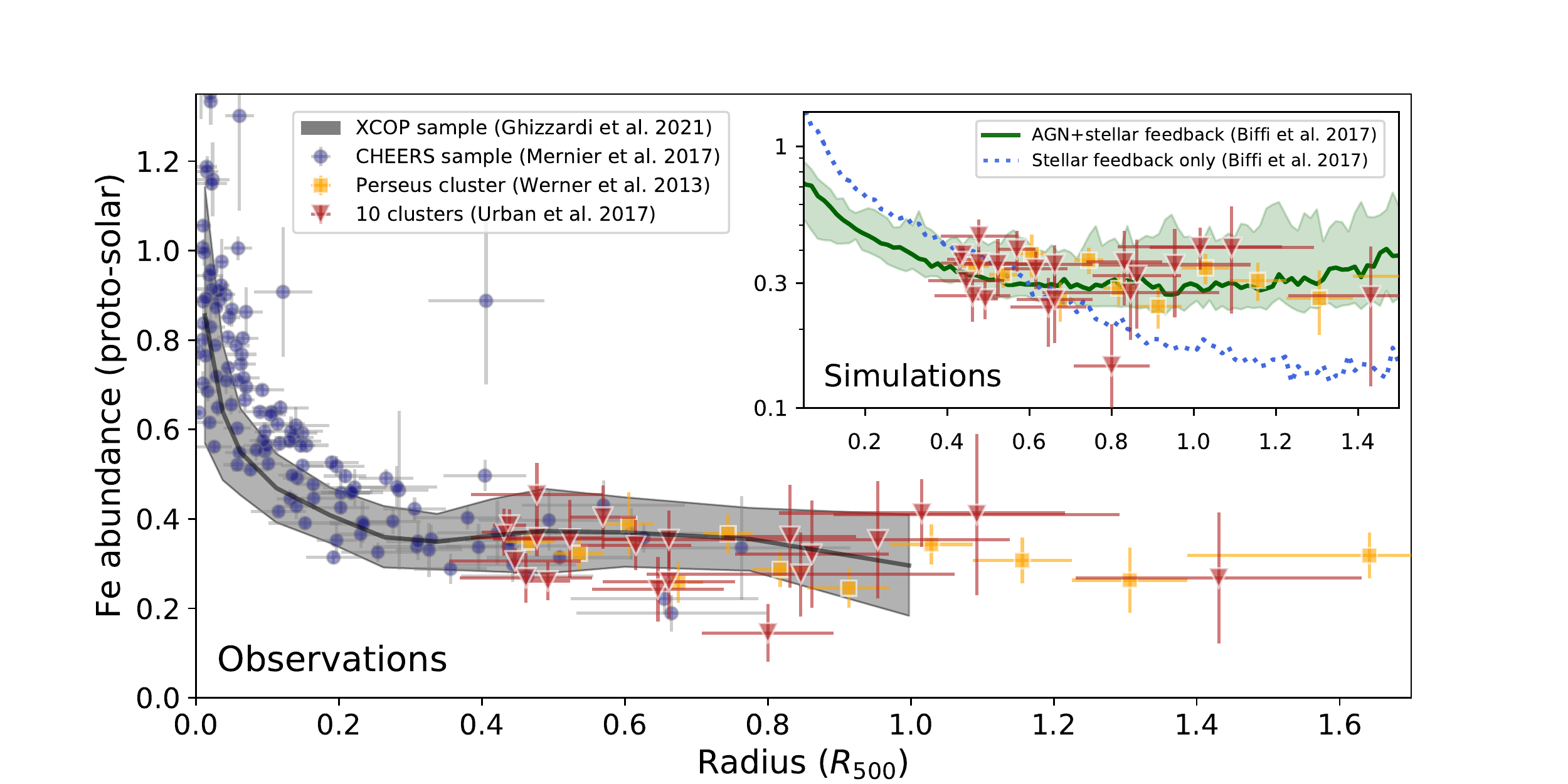}
    \caption{Compilation of radial profiles of Fe abundance in cool-core clusters from the literature. The main panel shows measurements from observations: the XCOP \citep{ghizzardi2021} and CHEERS samples \citep{mernier2017} using \textit{XMM-Newton}/EPIC; and measurements at cluster outskirts using \textit{Suzaku}/XIS \citep{werner2013,urban2017}. The inset panel compares the latter outskirts measurements with predictions from the Dianoga cosmolgical hydrodynamical simulations \citep{biffi2017}, assuming either solely stellar feedback or including AGN feedback as well. Abundances are all expressed in units of A09.
    }
    \label{fig:Fe_profs}
\end{figure}
Beyond the innermost regions, namely at radii $r \gtrsim 0.3$--$0.5\,R_{500}$, the radial metallicity profiles are observed to flatten and reach a uniform enrichment level around a value of $\sim 0.3$ solar (assuming recent reference solar values; e.g. by A09), out to $R_{200}$ and beyond. 
This extended earlier findings agree with \textit{BeppoSax} and \textit{XMM-Newton} which provided consistent results on the enrichment level of intermediate cluster regions~\cite[e.g.][]{degrandi2001,degrandi2004,leccardi2008}.
Nowadays, this remarkably uniform metallicity is systematically observed at intermediate and outer radii of nearby systems, in both cool-core and non-cool-core clusters.
Given the large ($\sim$Mpc) spatial scales covered by the outskirts, a central question arises about the origin of such spatially uniform metal distribution: can this distribution be explained by in-situ metal enrichment? If metal-rich gas stripped from present-time cluster galaxies were playing an important role for the enrichment, then abundance gradients would
\VB{match that of the galaxy distribution and thus}
be steeper and the whole metal distribution would be less homogeneous. 
\FMbis{Similarly, the active galactic nucleus (AGN) found within the central brightest cluster galaxy (BCG) can redistribute metals outside of the core only to some extent (Sect.~\ref{sec:when:transport}), while it \textit{cannot} provide the dominant process for spreading chemical elements all the way to the cluster outskirts. In fact, if the bulk of metal release would happen at late stage, the well stratified entropy of today's ICM would inhibit the mixing of metals and ``freeze'' their spatial distribution to some extent.}

Instead, the most  
\VB{accredited hypothesis}
is that the bulk of metals release from stars in galaxies into the ICM  has happened \textit{before} the cluster assembly, i.e. at redshifts $z\gtrsim 2$--$3$, coinciding with the main epoch of cosmic star formation \citep{madau2014}.

In Fig.~\ref{fig:Fe_profs} we show a compilation of observed radial profiles of Fe abundance in cool-core clusters taken from the literature~\citep{werner2013,mernier2017,urban2017,ghizzardi2021}. These indicate the decreasing gradient as a function of cluster-centric distance out to $\sim 0.3\,R_{500}$ and a flat trend in the outer regions till $\sim 1.5$--$1.6\,R_{500}\sim R_{200}$. In the top-right inset panel, a zoom onto the outskirts is reported, where we contrast the observational results against radial profiles of galaxy clusters extracted from the cosmological hydrodynamical simulations by~\citet{biffi2017}.
Investigations of simulated galaxy clusters, led by various independent groups, find indeed flat Fe profiles in the intermediate and outer regions of present-day systems, both cool-core and non-cool-core ones, in agreement with the above observed enrichment level of about $\sim 0.3$ solar \VB{with respect to the reference abundances by A09}~\cite[see][for a recent review]{biffi2018b}.
Simulations essentially confirm a remarkably uniform metallicity in cluster outskirts, within individual clusters but also from system to system, as confirmed by the small scatter on the median profiles (see e.g.\ Fig.~\ref{fig:Fe_profs}).
By including different feedback processes, simulations further allowed to investigate the \textit{origin} of this uniform enrichment.
Whereas the exact  
\VB{normalisation} of the profiles can be sensitive to the
choice of the metal yields and IMF adopted in the sub-grid chemical models embedded in the simulations, the overall shape of the profiles can be used to assess the role of different enriching mechanisms in shaping the resulting chemical pattern~\cite[][]{fabjan2010,biffi2017,vogelsberger2018}.
In particular, early simulations including only stellar feedback typically resulted into steeper metallicity gradients, that rapidly decrease from the centre outwards. In contrast, when also early AGN feedback is accounted for, the ICM metallicity profile of today's clusters is evidently flatter, especially at $\gtrsim 0.3\,R_{500}$, and better agrees with the observed trend -- as visible from the comparison reported in the inset panel of Fig.~\ref{fig:Fe_profs}. Interestingly enough, the $z \sim 2-3$ epoch 
\VB{when the bulk of the early enrichment must have happened}
also corresponds to the cosmic peak of supermassive back hole activity \citep{hickox2018}, further in line with the important AGN feedback effect predicted by simulations.
\VB{In this scenario, the central AGN is not the only one to play a role, but the final homogeneous metal distribution is the result of the cumulative effect, integrated over cosmic time, of all AGN in the large scale region ($\sim 10$\,Mpc) that collapses to form a present-day cluster.}

Summarising, the remarkably uniform metal distribution found in cluster outskirts by both observations and simulations strongly suggests a scenario in which metals were massively produced and released more than $\sim$10~billion years ago, i.e. before clusters started to form. The key actor in this \textbf{early-enrichment} scenario is definitely the strong feedback from supermassive black holes, which significantly helped to eject metals out of the galaxy host and to mix it homogeneously within the inter-galactic plasma that later formed the ICM. In the next subsections, we will see how other results from observations and/or simulations further support this scenario.

\subsection{Mechanisms for metal transport}\label{sec:when:transport}

\FMbis{As we have seen, feedback from supermassive black holes is thought to have played a considerable role in ejecting, stirring and mixing a very large fraction of metals outside of galaxies at early cosmic epochs. This \textbf{\textit{early} AGN feedback} has probably occurred in individual galaxies before clusters existed, when their respective AGNs were accreting gas at tremendous rates (leading to powerful winds referred to as the "radiative mode"; see Section XII: Active Galactic Nuclei in X and Gamma-rays).} 

\FMbis{In today's groups and clusters, AGN feedback is mostly seen through the supermassive black hole of the central BCG, accreting  material at much lower rates. This "kinetic" (or "radio") mode is known to trigger jets or lobes of relativistic plasma (visible in radio via synchrotron emission) which interact with the hot gas and shape cavities in the X-ray surface brightness images of clusters and groups (see Chapter: AGN feedback in groups and clusters). This \textbf{\textit{later}, central AGN feedback} is even thought to heat the central cooling gas, and somehow regulate its heating-cooling balance (see also Chapter: Thermodynamical profiles of clusters and groups, and their evolution). Quite interestingly, it has been observed that these jets may be efficient at displacing metals out of cluster/group cores. Using spatial spectroscopy to build metal maps, several studies had indeed measured a significant metal enhancement along radio jets \citep[e.g.][]{simionescu2008,kirkpatrick2009,kirkpatrick2011}.}

Besides \FMbis{AGN feedback}, other mechanisms also contribute to the transport of gas from the galaxies to the surrounding medium and to the mixing of metal-rich gas~\cite[][]{schindler2008}.
As mentioned before, also \textbf{stellar feedback} due to winds from supernovae and young massive stars can transfer significant amounts of energy to the surrounding gas, thus promoting \VB{galactic fountains and outflows}. This process can therefore transport gas and metals from the galaxies into the ICM, especially when star formation is active and massive stars or SNcc form and release chemical elements. 
While possibly contributing, cosmological simulations showed nevertheless that stellar feedback alone can account for the observed metallicity levels but typically produces too steep gradients, failing to match the uniform late-time metal distribution observed today in cluster outskirts.
\VB{Similarly to AGN feedback, stellar feedback can be more effective before the cluster assembles, i.e.\ at $z\gtrsim 2$. At low redshifts,} 
stellar winds can be suppressed by the pressure of the ICM especially in the innermost regions, at the bottom of the cluster potential well.

Nonetheless the ICM pressure onto cluster galaxies can also promote transport processes such as \textbf{ram pressure stripping}~\cite[][]{gunn1972}, through which gas gets efficiently removed from the galaxies and eventually mixed to the ICM~\cite[][]{boselli2006}.
This is supported by evidence of galaxy gas tails, for instance by HI observations~\cite[][]{smith2010}, and found in simulations as well.
Compared to galactic winds, stripping can be more efficient in transporting gas and metals from galaxies into the ICM, especially in massive clusters and at low redshifts, where the star formation rate decreases and the formed ICM has a stronger impact.
The dense environment in clusters is potentially suitable for \textbf{galaxy-galaxy interactions} as well. A close encounter between two cluster galaxies can favour star bursts, leading to galactic winds, or stripping and tidal interactions, with subsequent removal of metal-rich ISM from the galaxies.
\VB{Furthermore, \textbf{gas sloshing}\footnote{\FM{See also Chapter: The Merger Dynamics of the X-ray Emitting Plasma in Clusters of Galaxies.}} is able to spread heavy elements located close to the central galaxy from the centre outwards. This mechanism has been advocated to explain the cold fronts observed in the core of the Virgo cluster, where its role in transporting metals is suggested to be more significant than the interaction between the central AGN and the ICM~\cite[][]{simionescu2010}.}

Other than processes displacing gas from galaxies into the ICM, metals can be \VB{freshly produced and} mixed into the ICM by the \textbf{intracluster stellar component} as well \FM{\citep[for a review, see][]{contini2021}}. This population of stars, predicted by simulations and observed in real clusters, sums up to $20$--$50\%$ of the total stellar mass in galaxy clusters and can enrich the ICM directly, without the need of overcoming galactic potential wells.

\subsection{Galaxy clusters and groups: similar or different enrichment?}

\begin{figure}[t]
 \centering
     \includegraphics[width=0.95\textwidth]{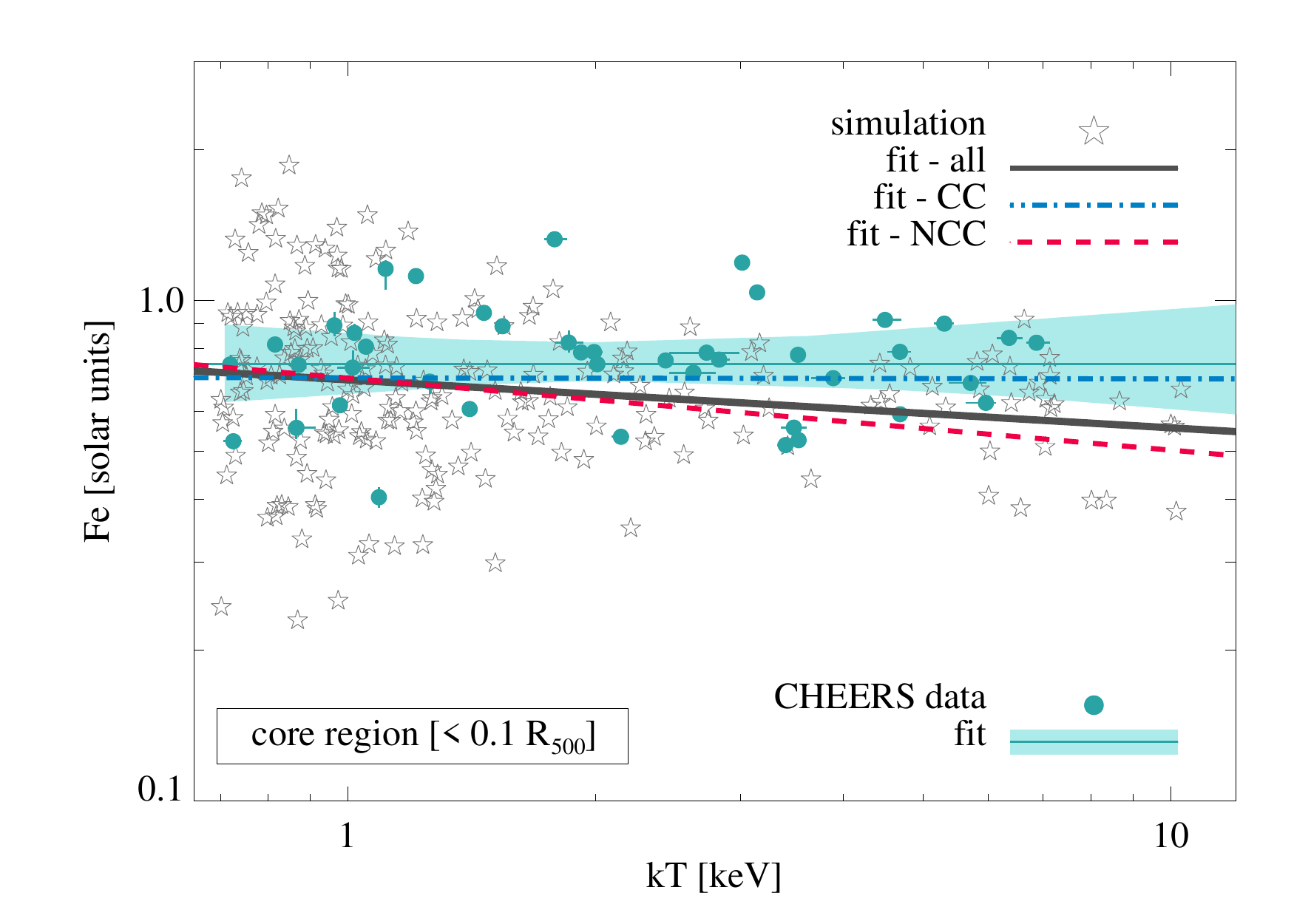}
     \caption{Comparison between average gas temperature in the core of clusters/groups (i.e. tracing their total masses) and their average Fe abundance. Observations (data and their best-fit trend) are shown in turquoise \citep{mernier2018a}. Simulated systems and their best-fit trends are shown by white stars and straight lines, respectively \citep{truong2019}. All abundances are rescaled to the reference solar values by~A09. Reprinted with permission from \citet{gastaldello2021}.
     \label{fig:kTvsFe}}
 \end{figure}

Besides spatial distribution, one may also wonder whether the metallicity (or,  
\VB{specifically,}
Fe abundance) is on average lower, similar, or higher in groups compared to clusters. An easy (yet robust) way to do such comparison is to plot the Fe abundance in the ICM of various systems as a function of their average temperature \FM{(thus their mass)}. In fact, as seen in \FM{Chapter: Scaling relations of clusters and groups, and their evolution} in this book, the temperature of a group/cluster is known to be an accurate tracer of its total (visible and dark matter) mass. Of course the physical size over groups and clusters is very different, and using a common fraction of $R_{500}$ is important in order to compare all these systems in a consistent way. Such a temperature-abundance comparison is shown in Fig.~\ref{fig:kTvsFe}, for both observed and simulated clusters. Simulations should be performed on a wide enough area to minimise sub-grid physics issues (modelling of AGN feedback, subtle dynamical effects, etc.) while, on the contrary, observations of cluster cores are more accurate (because they are less affected by background). A good compromise is found for $R < 0.1 R_{500}$, which is adopted here. 

The figure shows \FM{that clusters} remain on average as enriched as groups. This is particularly striking when comparing the best-fit line for simulated cool-core systems (dash-dotted blue) with the best-fit trend for observed data (turquoise envelope; from the CHEERS sample, all being cool-core systems too). \FM{The dispersion in Fe abundance (along the y-axis), due to different chemical histories for each system, is also consistent between observations and simulations. } 
Such similarity of global enrichment between groups and clusters is very important to note and brings another evidence toward the early-enrichment scenario. Finding as much Fe is indeed in line with the idea that metals were widely present and dispersed in the intergalactic medium in the proto-cluster/group region before the assembly of the galaxy system completed.

\subsection{Chemical composition of the ICM}\label{sec:when:ratios}

\begin{figure}[t]
 \centering
     \includegraphics[width=0.95\textwidth]{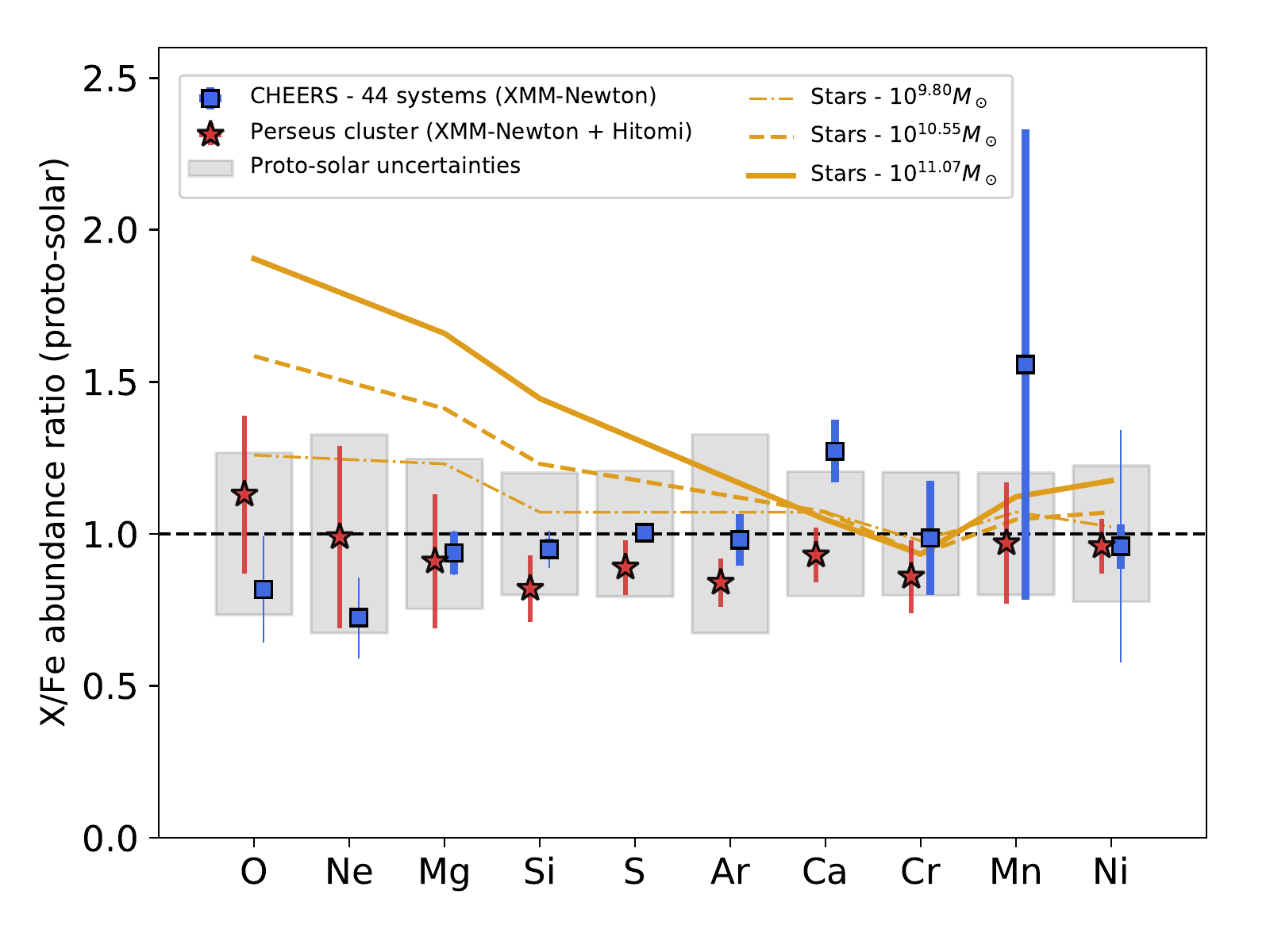}
     \caption{The chemical composition of the ICM. Abundance ratios (X/Fe) are measured from (i) a sample of 44 systems (with \textit{XMM-Newton} EPIC and RGS observations) \citep{mernier2018b} and (ii) the bright, nearby Perseus cluster (with the SXS micro-calorimeter onboard \textit{Hitomi} and \textit{XMM-Newton} RGS observations) \citep{simionescu2019}. For comparison, we also show stellar abundances measured in early type galaxies of three different mass ranges \citep{conroy2014}, as well as the typical proto-solar uncertainties (in proto-solar units of L09).}
     \label{fig:ratios}
 \end{figure}

Beyond Fe, the other chemical elements detected in the ICM are essential  
to better understand the history of its enrichment. In fact, measuring the average abundance \textit{ratios} X/Fe (i.e. typically given over the Fe abundance) of the ICM \FM{enables to directly compare its \textit{chemical composition} with that of many other astrophysical objects (stars, our Solar System, interstellar medium in galaxies, etc.)}. Two independent results are shown on Fig~\ref{fig:ratios}. The first one measured these ratios in the ICM of 44 clusters and groups (the CHEERS sample) using the moderate-resolution EPIC instruments (for Mg/Fe, Si/Fe, S/Fe, Ar/Fe, Ca/Fe, Cr/Fe, Mn/Fe, and Ni/Fe) and the RGS gratings (O/Fe and Ne/Fe) onboard \textit{XMM-Newton} \citep{mernier2018b}. The second one focused on a very bright and nearby system -- the Perseus cluster -- taking advantage of the exquisite spectral resolution offered for the very first time by the SXS instrument onboard the past mission \textit{Hitomi} (Si/Fe and heavier), as well as on the RGS gratings onboard \textit{XMM-Newton} (O/Fe, Ne/Fe, and Mg/Fe) \citep{simionescu2019}. Completing earlier signs, these two studies show that the chemical composition of the ICM is remarkably similar to the composition of our own Solar System (i.e. corresponding to the proto-solar units of L09).  This result is very surprising, as there is \textit{a priori} no reason for these two very different astrophysical  
\VB{environments}
to be chemically so similar. Indeed, our Solar System belongs to the Milky Way, a spiral galaxy \FM{of $\sim 6 \times 10^{10} M_\odot$ of stellar mass \citep{mcmillan2011}} that has experienced successive episodes of star formation -- hence, being constantly enriched by SNcc products. Galaxies in clusters, on the opposite, are often defined as ``red-and-dead'' -- i.e. almost exclusively composed of old, red, low-mass stars. Whereas these cluster galaxies often exhibit super-solar \FMbis{abundance ratios} of their stars\footnote{This effect, that is more pronounced in more massive cluster galaxies, is attributed to the fact that these stars have formed early on and very rapidly (a scenario sometimes referred to as ``downsizing''). At the time of their formation ($z \gtrsim 3$), these proto-stars were accreting an interstellar medium that was still very young, already enriched by previous populations of SNcc but \FM{not} yet polluted by SNIa explosions. These early-formed stars have thus ``fossilised'' this very early chemical state of formation of their galaxy host.} (Fig.~\ref{fig:ratios}), \FMbis{their stellar populations have mostly low masses ($< 8~M_\odot$) and long lifetimes, implying} that they should in principle explode as SNIa and inject large amounts of Fe-peak elements in the ICM over cosmic time. Here again, the early-enrichment scenario could explain this puzzle: if stars and supernovae had completed the enrichment of their surrounding gas during or before cluster formation (hence, before their galaxies became red-and-dead), one can expect comparable amounts of SNIa \textit{and} SNcc products to end up in the (still forming) ICM. Pushing this further, one could speculate that this ``proto-solar'' composition corresponds to a rather ``universal'' value, typical of the gaseous content of our Universe (i.e. most of the baryons). This gas, pre-enriched at this level, could then eventually collapse into localised regions of galaxy discs such as in our Milky Way. These solar abundance ratios are more difficult to predict directly from simulations, essentially because they highly depend on the metal yields that are assumed in simulated clusters. On the contrary, the observed ratios are useful to constrain yields, which can be in turn adopted in simulations to improve the enrichment modelling and obtain ICM chemical properties in better agreement with observations 
(see also Sect.~\ref{sec:stellar_physics}). 

Not shown on this figure is the N/Fe abundance ratio, which has been determined in the hot atmosphere of a few massive galaxies \citep[e.g.][]{mao2019} using \textit{XMM-Newton}/RGS. In fact, the dispersive nature of the instrument allows to detect the N line in a handful of compact systems only. Unlike the other ratios, N/Fe seems to deviate from proto-solar ratios, with clearly higher measurements ($\sim$2 proto-solar in units of L09). This deviation strongly suggests that the enrichment of hot atmospheres via AGB stars, the main source of N production, is decoupled from the supernovae (early) enrichment and may be still ongoing. Similarly, the C/Fe ratio also offers the potential to better understand the ICM enrichment through AGB stars. The only C emission line (C VI) shining in the X-ray band, however, is often very challenging to detect, even using \textit{XMM-Newton}/RGS over bright systems \citep{werner2006}. Future observatories will definitely help to constrain this ratio more accurately \citep{mernier2020} (see Sect.~\ref{sec:future}).

\begin{figure}[t]
 \centering
     \includegraphics[width=0.95\textwidth]{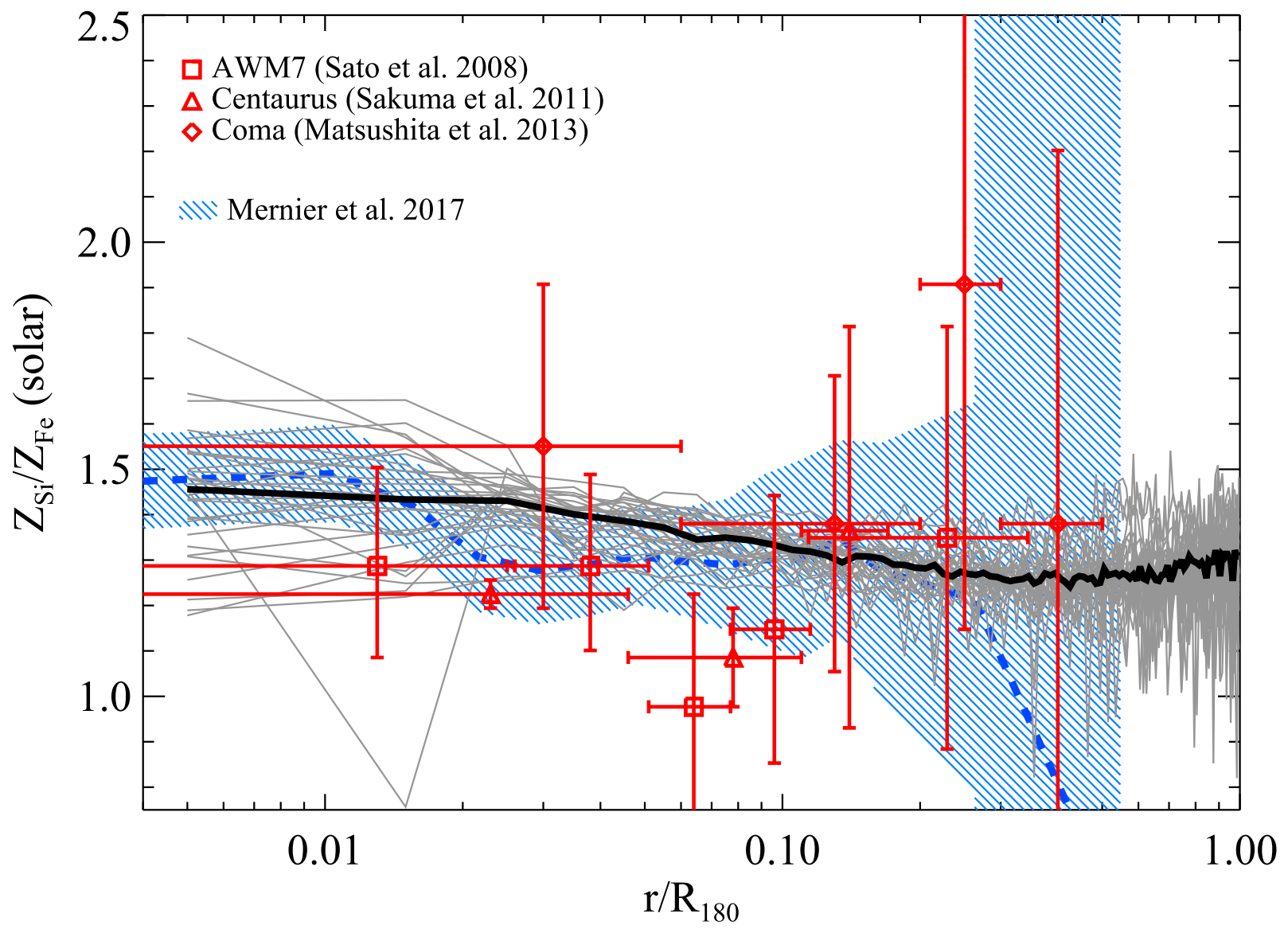}
     \caption{Variation of the Si/Fe ratio across radius. Red data points and the blue envelope are taken observational data. The black line and grey envelope are predictions from simulations. All values are rescaled with respect to the reference solar values of AG89. Reprinted with permission from \citet{biffi2018b}.}
     \label{fig:radial_ratios}
 \end{figure}

As we have seen in Sect.~\ref{sec:when:profiles}, valuable information on clusters chemical history reside not only in their global measurements, but also (and especially) in their spatial distribution. In the context of the chemical composition of the ICM, one can therefore go one step further by investigating its \textit{radial distribution} as well. In other words: how do the X/Fe ratios -- and particularly the $\alpha$/Fe ratios -- vary from the core to the outskirts? This question is essential here to track the \textit{differential} chemical history of clusters, i.e. the individual evolution of the SNcc and SNIa enrichment channels. For quite some time, it was thought that the central metal peak in cool-core clusters was specific to SNIa only: as the central BCG is red-and-dead, the only sources of Fe at present times are SNIa. And as these SNIa explode often with a large delay after formation of their low-mass progenitors, the Fe peak was naturally associated to a recent central enrichment by SNIa. On the contrary, SNcc are not expected to occur in today's clusters anymore, as (almost) no massive stars are seen in cluster galaxies. This means that, unlike Fe, the radial abundance profile of $\alpha$-elements was expected to be flat everywhere, hence the $\alpha$/Fe ratios should increase with radius. With time, however, measurements from observations became more accurate, and eventually revealed a central peak for \textit{all} measurable elements -- including O, Mg, and Si. This translates into no spatial variation of X/Fe ratios: the ICM chemical composition remains the same in cluster core and outskirts. This uniformity is also seen in cosmological simulations, as shown in Fig.~\ref{fig:radial_ratios}. Since, again,  
\VB{there are no evidence of recent stellar production in BCG and thus of short-lived massive stars,}
the central peak of $\alpha$-elements  
\VB{is due to}
an enrichment that took place early in cluster history, with \textit{no} recent contribution from  
\VB{SNcc.} \FM{By extension, since the central Fe peak follows remarkably well that of $\alpha$-elements, a similar early-enrichment scenario for SNIa comes as a natural explanation even within the central BCG. Unless \FMbis{there is a} remarkable coincidence in their spatial distribution, it is very unlikely that it would originate from recently exploded SNIa. The question of what happens with metals freshly produced by SNIa (i.e. from the low-mass star population) in the BCG is still debated; however residual star formation suggests that at least part of these metals may be locked into new stars.}

\subsection{Metal budget in clusters}

As mentioned earlier, clusters have a deep potential well. This means that the gas accreted during the cluster formation history will, in principle, never escape back outside of its zone of influence. The same reasoning can be naturally applied to metals: all chemical elements created by stellar populations inside a cluster should remain inside the cluster -- should it be in the form of ions wandering in the ICM or simply recycled back into new generations of stars in cluster galaxies. An interesting exercise is thus to estimate the total amount of metals in a given cluster -- i.e. locked in stars \textit{and} released in the ICM -- and to compare it with the amount of metals \textit{expected} to be reasonably produced by the whole stellar population in that cluster. 

In practice, and taking the case of Fe for convenience, one can simply look at the so-called \textit{iron yield}, defined as the ratio of the total mass of Fe (i.e. in the ICM \textit{and} in stars) observed in a given system over the total initial mass of the stars in that system (i.e. also accounting for the matter initially belonging to the stars that had been lost or ejected during their lifetime). Another quantity sometimes considered in the literature is the \textit{iron mass-to-light ratio},  
obtained by dividing the Fe mass from the ICM (easily derived from the Fe abundance and from the enclosed gas mass, \FMbis{the latter of} which can be obtained from \FM{X-ray} spectroscopy or even \FM{X-ray} surface brightness only) by the K-band luminosity at optical wavelengths (known to scale well with the stellar mass of the considered system). When comparing the above defined quantities with theoretical expectations (considering e.g. the fraction of stars ending up in SNIa or SNcc, as well as the Fe yields predicted for these supernovae), it appears that the amount of metals present in clusters is 2 to 10 times larger than what their stars could have reasonably produced \citep{loewenstein2013,renzini2014}. Interestingly, this problem seems to be alleviated for galaxy groups: as we have seen they have a similar metallicity as for clusters, however they retain less hot gas in comparison (i.e. their gas fraction is lower than for clusters). Given that significant amounts of gas -- hence of metals -- must have managed to escape the shallower gravitational well of groups, the better agreement between observed and predicted metal budget at group scales does not necessarily guarantee 
\FMbis{that our understanding of metal enrichment in groups is complete} \citep{gastaldello2021}.

Regardless of clusters or groups, however, this metal conundrum should be posed with caution. Many uncertainties  
\VB{still remain,}
e.g., the metal yields and the fraction of stars ending up in supernovae. In any case, this seems to suggest that the stellar population observed in clusters today has 
\FMbis{little} to do with metals that are found in their ICM. This may be seen as another argument in favour of (or, at the very least, not contradicting with) the idea that metals were released by early stellar populations that enriched their circum-galactic medium before galaxies assembled into clusters \citep{blackwell2021}. Dedicated cosmological simulations on this question will be helpful to better understand the (stellar and gas) metal budget in clusters and groups.

\subsection{Redshift evolution of the chemical enrichment}

\begin{figure}
    \centering
    \includegraphics[width=1.0\textwidth, trim={0cm 0cm 0cm 0cm},clip]{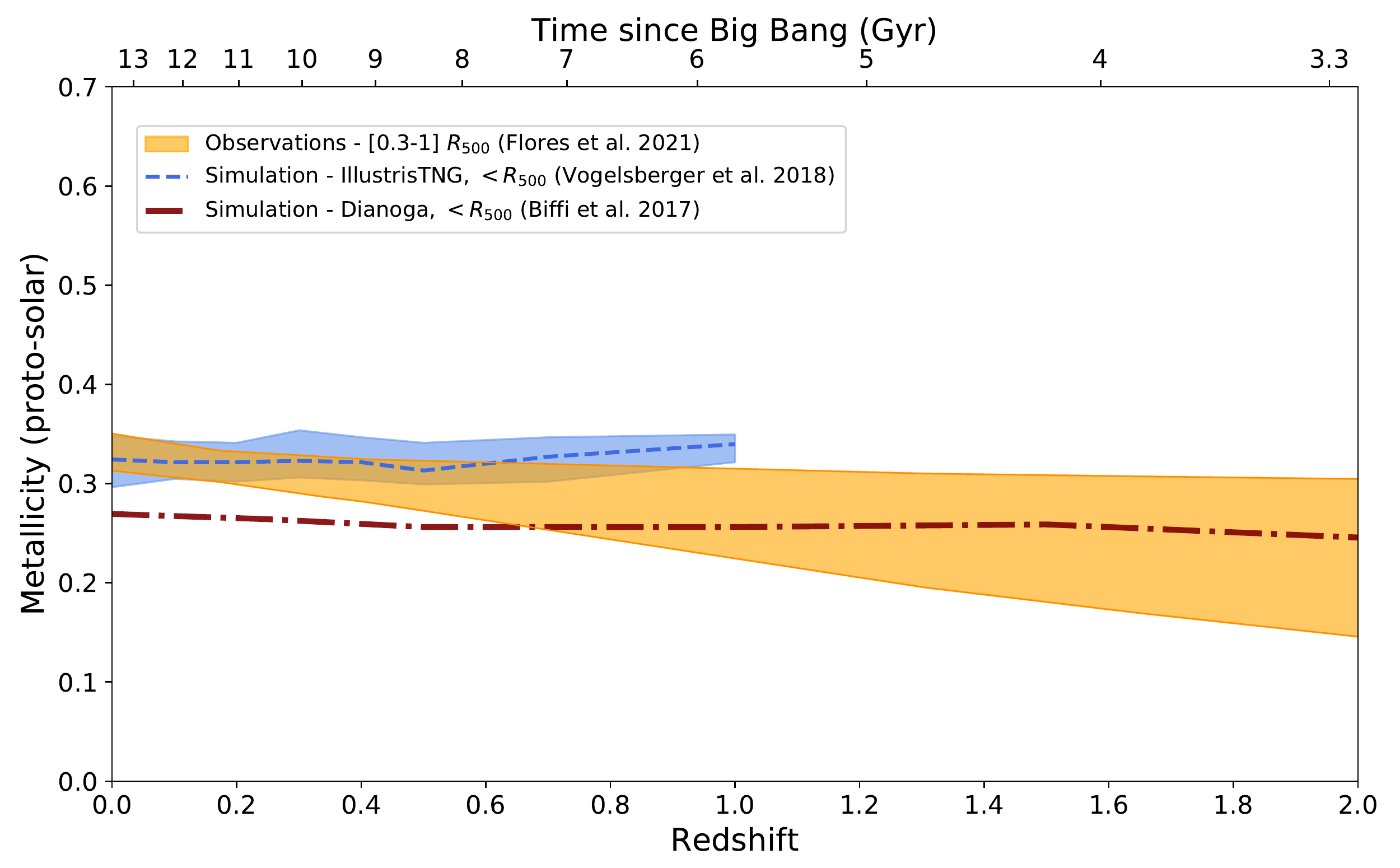}
    \caption{Redshift evolution of cluster metallicities, as seen by recent observations \citep{flores2021} and as predicted by cosmological simulations -- Dianoga \citep{biffi2017} and IllustrisTNG \citep{vogelsberger2018}. Metallicities are all expressed in units of A09.
    }
    \label{fig:Fe_evol}
\end{figure}

So far we have discussed how the uniform distribution of metallicity and abundance ratios out to large radii in local galaxy clusters and the consistent enrichment level from clusters to groups point towards an early enrichment of the ICM. Although extremely promising, all these evidence remain indirect. As undoubtedly a holy grail in this field, a key accomplishment would be to \textit{directly} track back the ICM metallicity in clusters up to high redshifts, in order to find evidence of its (non-) evolution with cosmic time.
\FM{Such observational attempts at high redshift are still very challenging, essentially because they are conducted on very small, hand picked samples. Moreover, global metallicities are derived mainly from the Fe-K line emission, which is difficult to detect with enough statistics in distant sources. Nevertheless,} in the last two decades several observational studies have been dedicated to measuring the average abundance in the core and outer regions of galaxy clusters out to $z\sim1$ or more~\cite[][]{mernier2018c}.
Findings generally agree on a very mild evolution of the ICM metallicity below redshift $z\sim 1$--$1.3$, which mostly concerns the innermost region and possibly extends to the intermediate radii~\cite[][]{ettori2015,mcdonald2016,mantz2017,liu2020}. At larger cluster-centric distances the enrichment level is observed to be nearly constant with time.
Recent investigations by~\citet{flores2021} of deep \textit{XMM-Newton} observations of 10 massive galaxy clusters at redshifts $1.05 < z < 1.71$ further confirmed this picture. They reported an average metallicity in the $0.3 < r/R_{500} < 1$ radial range, about $0.21\pm 0.09$ solar (in units of~A09), which is consistent, albeit slightly lower than, the average outskirt metallicity of low-redshift clusters.
Cosmological simulations provide similar results, also finding cluster metallicities that are consistent with no evolution at $z\lesssim 2$, especially in the outskirts~\cite[][]{biffi2017,vogelsberger2018,pearce2021}.
This is visible in Fig.~\ref{fig:Fe_evol}, which reports the results from cosmological simulations by~\citet{biffi2017} and~\citet{vogelsberger2018} for the global cluster region within $R_{500}$, compared with the compilation of observational data presented by~\citet{flores2021} for the cluster outskirts, at $0<z\lesssim2$.
The core enrichment level also shows no significant evolution in simulated clusters, suggesting that the mild trend observed in real systems could rather be driven by the evolution of the cool-core population, as discussed by~\citet[][]{ettori2015}. 
Here again, this measured and predicted absence of (or very minor) evolution at large cluster-centric distances points towards an early-enrichment (or even a pre-enrichment) scenario.
This scenario successfully explains both the chemical properties of clusters at low redshifts and the negligible evolution of the ICM metallicity below redshift $z\sim 2$.

\section{\FM{Understanding stellar physics from metals in the ICM}}\label{sec:stellar_physics}

We have seen how measuring accurate abundances in the ICM and comparing these measurements to state-of-the-art cosmological simulations provides invaluable information on the history and the evolution of the enrichment at the largest scales of the Universe. However, these accurate measurements have the potential to teach us even more than the ICM enrichment itself. 

Looking back at Fig.~\ref{fig:yields}, it is clear that the currently competing SNcc and SNIa models predict different abundance ratios. This means that, if one can accurately measure specific abundance ratios from a given supernova remnant, one should be able to constrain the physics and/or the environment of its explosion. This holds for SNIa, whose progenitors are still not well understood (single- v.s double-degenerate scenario), as well as for SNcc, from which the \FMbis{initial mass and} metallicity of their \FMbis{progenitor star} can be constrained (Sect.~\ref{sec:SNe}). In practice, however, supernova remnants are complicated objects. Like a snowplow, metals produced by supernovae and propagated near the shock of its explosion may sometimes get mixed with metals accumulated from the surrounding interstellar material that was already in place before the explosion. \FM{Similarly, SNe exploding in dense regions might have occurred as a chain from different progenitors.} Moreover, the ionisation state of the expanding gas is also very complex 
(as its ionisation rate usually keeps evolving with time) and radiative transfer effects may occur -- meaning that the line equivalent widths in the X-ray spectra of supernova remnants are difficult to translate into chemical abundances. Lastly, an additional problem  
is simply that \textit{one} remnant is certainly not representative of the bulk of supernovae in the Universe. Extended further, even the few dozens of currently accessible supernova remnants in our Galaxy are clearly not enough to study these issues 
over a statistically meaningful sample.

At much larger scales, the ICM naturally solves all these problems. First, because as we have seen, the ICM is mostly optically thin and in collisional ionisation equilibrium, resulting into robust abundance measurements from line equivalent widths. Second, because the presence of chemical elements in the ICM is the result of the explosion of \textit{billions} of supernovae through cosmic ages \FM{\citep[though, as seen before, the bulk of metals has been probably produced and released in the 3 first billion years of the Universe; e.g.][]{biffi2017,urban2017}}. Measuring accurate abundance ratios in the ICM (via deep observations) thus constitutes an easy and unique opportunity to better understand SNcc, SNIa, their explosions and/or their progenitors.

This concept translates very easily in practice. The idea is to compare the X/Fe abundance ratios measured in the ICM of either one given cluster/group, or (even better) a sample of systems, with a set of combined SNcc+SNIa yield models available in the literature. The conversion from yields predicted by a given supernova model (metal mass of a given element, $M_\mathrm{X}$, usually in units of solar masses) to their associated predicted abundance ratios, $(\mathrm{X/Fe})_\mathrm{pred}$, can be easily done by noting that 
\begin{equation}\label{eq:yields}
(\mathrm{X/Fe})_\mathrm{pred} = \frac{M_\mathrm{X}}{M_{\mathrm{Fe}}} \, \frac{\mu_\mathrm{Fe}}{\mu_\mathrm{X}} \, \FMbis{\frac{N_{\mathrm{Fe},\odot}}{N_{\mathrm{X},\odot}}},
\end{equation}
where $\mu_\mathrm{X}$ and $\mu_\mathrm{Fe}$ are the atomic weights of elements X and Fe (in the same units as $M_\mathrm{X}$ and $M_\mathrm{Fe}$).

This equation holds for both types of supernovae.
Note, however, that SNcc yields are usually given \FMbis{as a function of} initial metallicity $Z_\mathrm{init}$ \textit{and} \FMbis{of} initial mass of the star progenitor: $M_\mathrm{X}(Z_\mathrm{init},m)$. To compare these SNcc models to X/Fe ratios in a consistent way, one first needs to integrate them over an assumed IMF:
\begin{eqnarray}
M^{\mathrm{int}}_{X}(Z_\mathrm{init}) &=& \frac{\int_{M_\mathrm{low}}^{M_\mathrm{up}} M_{X} (Z_\mathrm{init},m) \ \phi(m) \ dm}{\int_{M_\mathrm{low}}^{M_\mathrm{up}} \phi(m) \ dm}, 
\end{eqnarray}
\FMbis{where $\phi(m)$ can be parametrised as e.g. $\propto m^{-2.35}$ to adopt a Salpeter IMF (see Sect.~\ref{sec:techniques:sim})}
and $M_\mathrm{low}$ and $M_\mathrm{up}$ are the assumed stellar mass limits below and beyond which stars do not explode as 
SNcc\footnote{The exact value of these limits are still being debated in the literature, however they are often assumed as $M_\mathrm{low} = 10~M_\odot$ and $M_\mathrm{up} = 40~M_\odot$.}. This IMF-integrated mass yield $M^{\mathrm{int}}_{X}(Z_\mathrm{init})$, depending now only on the initial progenitor metallicity, can be injected into Eq.~(\ref{eq:yields}). Finally the total predicted ratios for the SNcc+SNIa models,
\begin{equation}
(\mathrm{X/Fe})_\mathrm{pred}^\mathrm{tot} = a(\mathrm{X/Fe})_\mathrm{pred}^\mathrm{SNcc} + b(\mathrm{X/Fe})_\mathrm{pred}^\mathrm{SNIa}
\end{equation}
with $a$ and $b$ as adjustable parameters, can be fitted to the observed ratios $(\mathrm{X/Fe})_\mathrm{obs}$ via a standard reduced $\chi^2$ method. These two parameters are not independent, as the case X = Fe imposes $a + b = 1$. Therefore, the fit depends only on one free parameter, often given as $ f_\mathrm{Ia} = b/(a+b)$; i.e. the relative fraction of SNIa (over the total number of SNe) having contributed to the ICM enrichment. Of course this ratio has to be interpreted with caution, as it does not necessarily reflect the true relative fraction of exploding SNIa in the nearby Universe. Indeed, depending on their environment, SNIa and SNcc may have specific (in)efficiencies to release metals outside of their galaxy hosts.

\begin{figure}[t]
 \centering
     \includegraphics[width=1.0\textwidth]{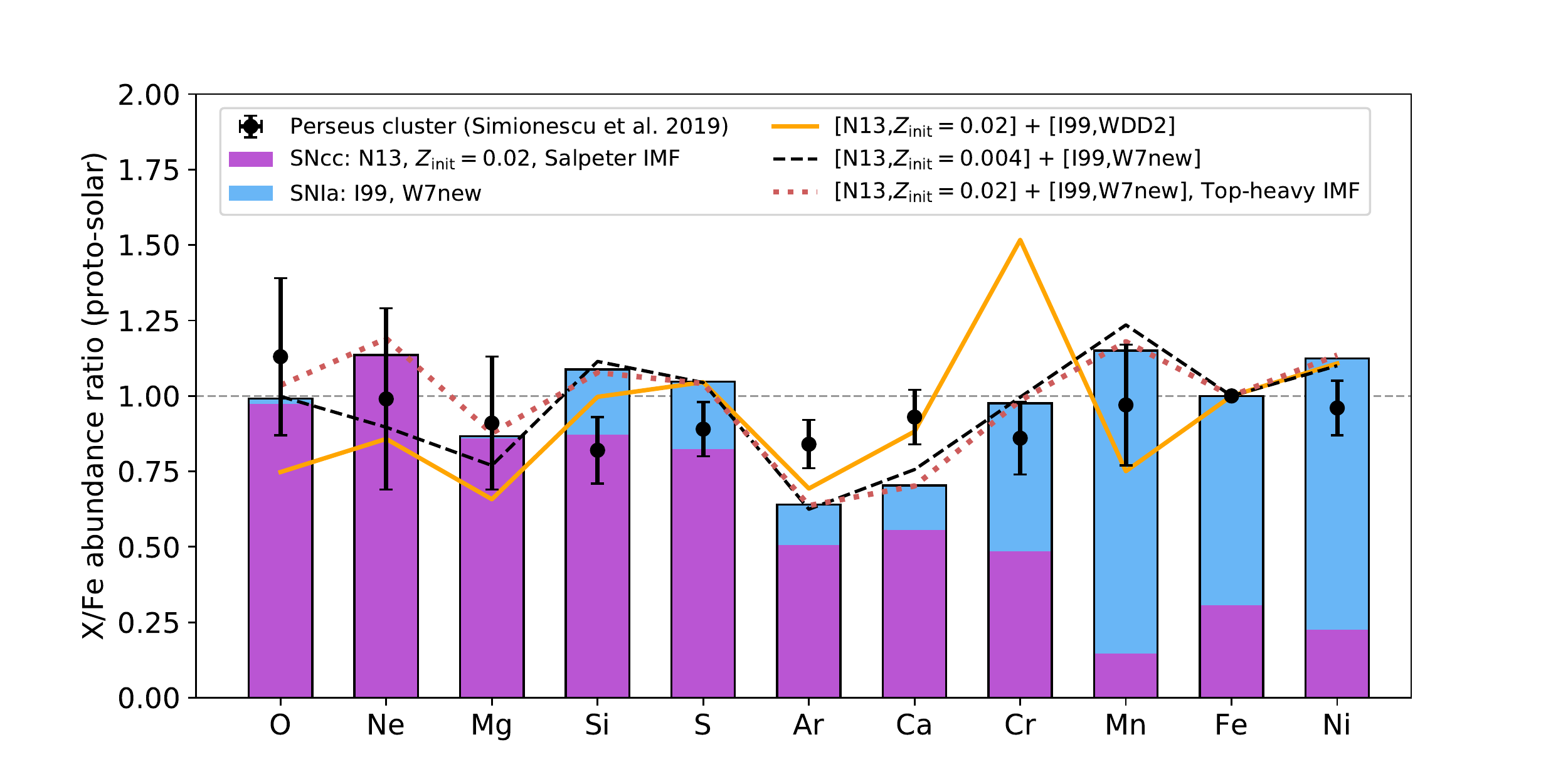}
     \caption{Abundance ratios measured in the core ICM of the Perseus cluster with \textit{XMM-Newton}/RGS and \textit{Hitomi}/SXS \citep{simionescu2019}, compared with various sets of theoretical SNcc+SNIa yield models. The best-fit set of models ($Z_\mathrm{init} = 0.02$ for SNcc integrated over a Salpeter IMF, together with an updated W7 deflagration model for SNIa, respectively from \citet{nomoto2013} and \citet{nomoto2018}), are shown with histograms, from which contribution from SNcc and SNIa is shown separately for each element. The total (SNcc+SNIa) contribution of other, less consistent models is shown by solid, dash, and dotted lines. In each of these additional sets of models, we have changed either the initial metallicity of the SNcc, or the explosion mechanism of the SNIa model (the WDD2 model is delayed-detonation), or the shame of the IMF on which the SNcc model is integrated. All the measured and modelled ratios are expressed in proto-solar units of L09. Figure partly adapted from \citet{simionescu2019}.}
     \label{fig:fit_ratios}
 \end{figure}
 
Figure~\ref{fig:fit_ratios} illustrates the above exercise with the most accurate X/Fe ratios measured in a cluster so far: the Perseus cluster using \textit{XMM-Newton}/RGS and \textit{Hitomi}/SXS data \citep{simionescu2019} (see also Sect.~\ref{sec:when:ratios} and Fig.~\ref{fig:ratios}). Several sets of SNcc+SNIa models are fitted to these observed ratios. Assuming a Salpeter IMF, the best fit 
is obtained for a combination of a SNcc model from \citet{nomoto2013} with $Z_\mathrm{init} = 0.02$ (i.e. close to the metallicity of our Sun)  
with a deflagration SNIa model (the W7 model from \citet{iwamoto1999} with updated electron capture and nuclear reaction rates). For this best-fit model, we show the relative contribution from SNcc and SNIa for each element. One naturally finds back the initial statement of $\alpha$-elements being mainly produced by SNcc while Fe-peak elements being mainly produced by SNIa. For comparison, we also show the total (i.e. SNcc+SNIa) contribution predicted from three other sets of models, though with larger $\chi^2$ value (i.e. less good fit quality). In each of these additional sets of models, one assumption is modified: (i) the SNIa deflagration (W7new) model is replaced by a delayed-detonation model (WDD2), (ii) the initial metallicity of the SNcc model is changed from 0.02 to 0.004, or (iii) the initial models are adopted, but this time assuming a top-heavy IMF ($\phi(m) \propto m^{-1.95}$). The $f_\mathrm{Ia}$ fraction is then typically found between 15\% and 40\%. This whole exercise shows the considerable importance of a few key ratios for further constraining the models: in particular Cr/Fe and Mn/Fe for the SNIa models, as well as \FMbis{O/Ne} for the SNcc models. Note that accurate ratios may even allow to disentangle single-degenerate vs. double-degenerate SNIa models, hence help to reveal the true nature of SNIa progenitors.

On the other hand, it also clearly appears that none of the current models is able to reproduce all ratios at once. For instance, all sets of models predict systematically too much S/Fe while not enough Ar/Fe (in other words, the S/Ar ratio is systematically overestimated). This suggests that current nucleosynthesis models 
still require further improvements. 
Tight synergies between more accurate measurements and more realistic models are the key to better understand how supernovae explode, and what is the nature and the environment of their progenitors.

\FM{Last but not least, accurate measurements of the N abundance (or its N/Fe ratio) have the potential to extend the above constraints to that of AGB stars. Currently, such measurements are difficult (though not impossible; e.g. \citealt{mao2019}) because the main N emission line can be well resolved with the RGS instrument onboard \textit{XMM-Newton} only. However, the high-resolution spectroscopy allowed by future missions will 
\VB{help putting} tighter constraints on its abundance, which will translate into invaluable constraints on the average initial metallicity and, \VB{possibly}, the IMF of the AGB population of clusters and massive galaxies.}


\section{Future prospects}\label{sec:future}

Throughout this Chapter we have seen an overview of how and when metals, initially synthesised 
\VB{within} stars and supernovae, get ejected outside of their galaxies and enrich the hot gas pervading galaxy clusters and groups.  
\VB{The} impressive progress made in X-ray observatories and numerical simulations over the last two decades have allowed to sketch a  
\VB{comprehensive} picture of this large-scale enrichment. However, a full understanding of the history and mechanisms of this enrichment will be  
possible only via further improvements at both observational and numerical levels.

\begin{figure}[!]
 \centering
     \includegraphics[width=0.95\textwidth]{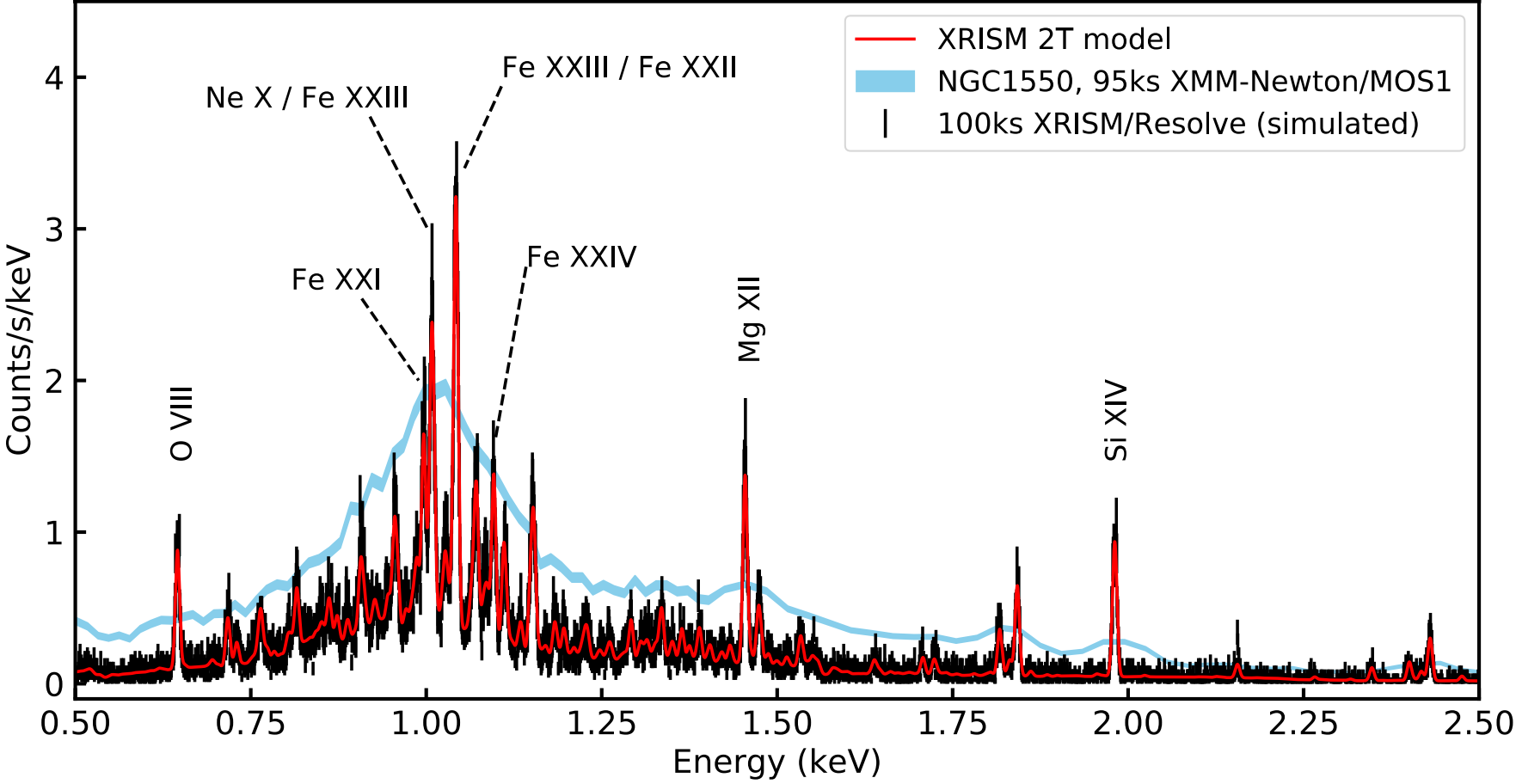}
     \caption{Comparison between a current \textit{XMM-Newton}/MOS spectrum (95~ks) and a simulated \textit{XRISM}/Resolve spectrum (100~ks, based on the best-fit values of the real MOS spectrum) of the hot gas of the early-type galaxy NGC\,1550. Reprinted with permission from \citet{gastaldello2021}.}
     \label{fig:xrism}
 \end{figure}

\begin{figure}[!]
 \centering
     \includegraphics[width=0.76\textwidth]{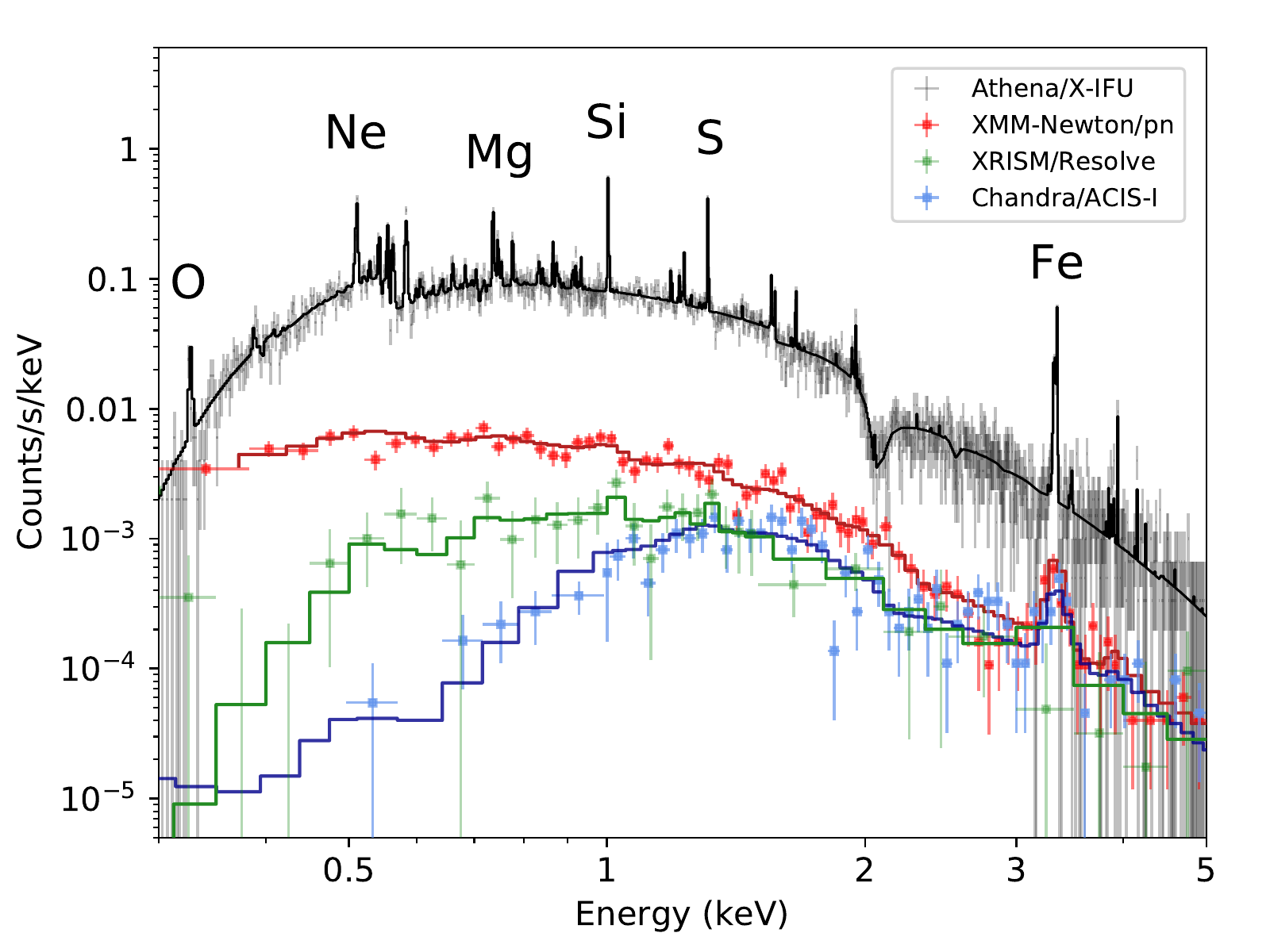} \\
     \includegraphics[width=0.70\textwidth]{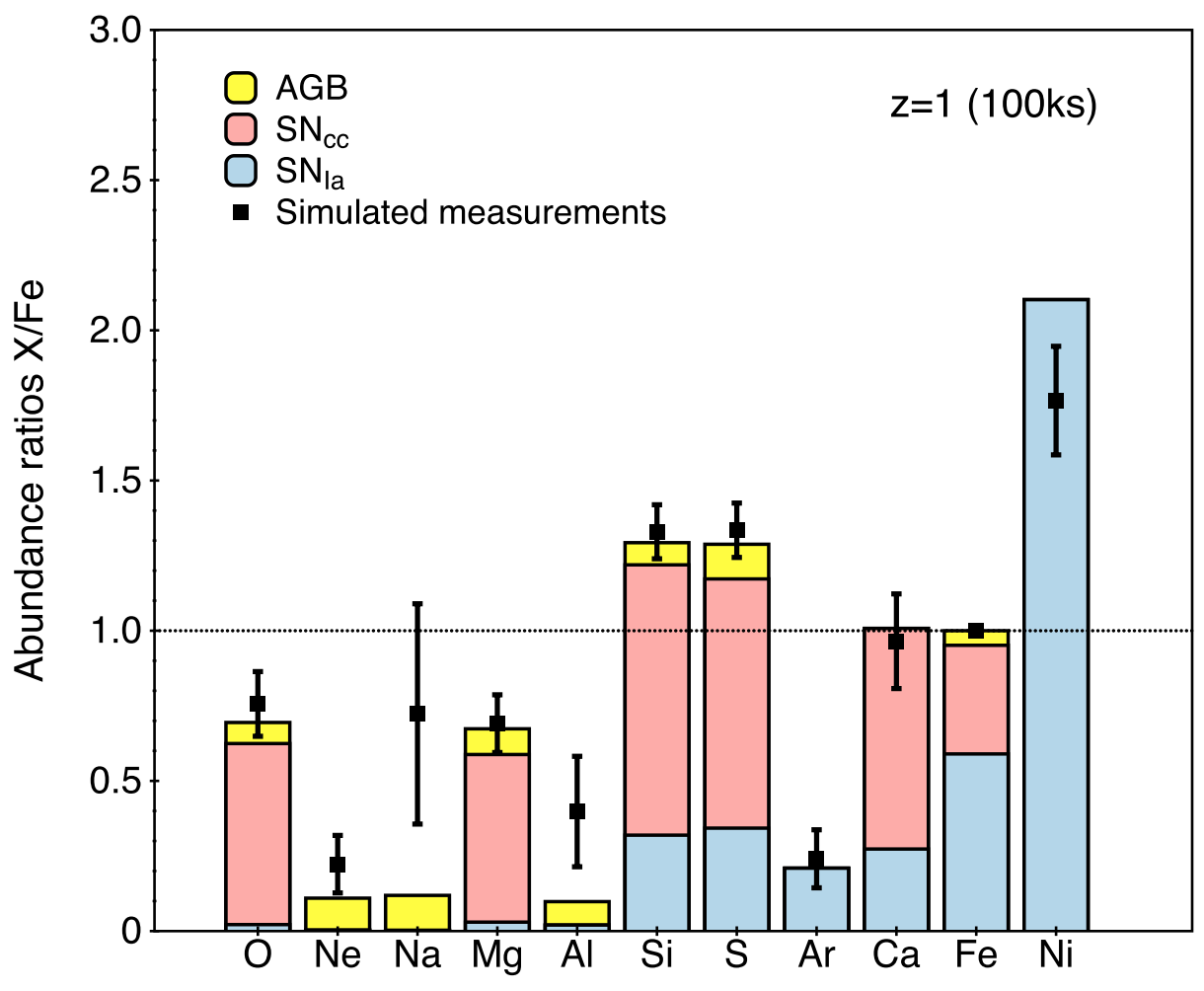}
     \caption{\FM{\textit{Top:}} Simulated spectra for a distant cluster at $z = 1$ ($kT$ and the abundances are assumed to be, respectively 3~keV and proto-solar in units of L09) with the \textit{Athena}/X-IFU, \textit{XMM-Newton}/pn, \textit{XRISM/Resolve}, and \textit{Chandra}/ACIS-I instruments. Adapted and reprinted with permission from \citet{mernier2018c}. \FM{\textit{Bottom:} Measured X/Fe ratios (error bars) from a mock \textit{Athena}/X-IFU observation of a simulated cluster at $z=1$, assuming specific AGB, SNcc and SNIa yields (histograms). Reprinted with permission from \citet{cucchetti2018}.}}
     \label{fig:athena}
 \end{figure}

On the observational side, \VB{as discussed above,} the current limitations are:  
(i) the spectral resolution, critical to probe in detail the enrichment level of  
nearby systems (ii) the effective area, critical to constrain precisely the abundances  
\VB{in} high-redshift systems,  
(iii) the spectral codes, critical to perform all these measurements \textit{accurately} (i.e. with negligible systematic bias). 
\VB{The first point will} be considerably improved \VB{in the next future}, thanks to the new era of \FMbis{X-ray} micro-calorimeters which will equip  
\VB{upcoming} missions. A first step will be achieved in 2023 by \textit{XRISM} -- a Japanese mission (with European and American contributions), and its micro-calorimeter instrument Resolve (5\,eV resolution, i.e. $\sim$24 times better than our current CCD detectors\FM{; see Chapter: XRISM}) \citep{tashiro2020}. This mission is essentially a re-flight of \textit{Hitomi}, which has already demonstrated the full power of high-resolution spectroscopy. An example is shown in Fig.~\ref{fig:xrism}, which further illustrates the dramatic magnification of spectral information from current CCD-like spectrometers (here \textit{XMM-Newton}/MOS) to \textit{Resolve}. In turn, resolving many more lines (especially in the still poorly resolved Fe-L complex) will unavoidably reveal a need for updating specific atomic transitions, which will lead to further revision of our spectral codes. This step will be very important, as accurate spectral codes are the key to measuring robust abundances for every chemical element in every system. Finally, the effective area will be considerably improved thanks to the next European mission \textit{Athena}.  
\VB{\textit{Athena},} to be launched in $\sim$2034 (with American \FM{and Japanese} contributions), will incorporate optics specifically designed to increase its effective area by almost one order of magnitude compared to \FM{\textit{XMM-Newton} -- our most sensitive X-ray observatory so far}.  
\VB{It} will have two instruments: the X-ray Integral Field Unit (X-IFU) \citep{barret2018} and the Wide Field Imager (WFI) \citep{rau2016}. The former consists  of 3168 micro-calorimeters (with \FM{2.5}\,eV resolution) forming an entire pixel grid covering a field of view of 5 arcmin equivalent diameter. The latter, although using a standard CCD-like detector, will benefit from a 40$\times$40 arcmin$^2$ field of view \FMbis{(see Chapter: ATHENA)}. Both instruments are thus  
complementary and will allow extraordinary progress at all levels: unprecedently accurate abundance measurements in clusters \textit{and} groups, spatial distribution of metals, and last but not least, the redshift evolution of multiple abundances, allowing to probe directly the enrichment history in clusters out to (and probably even beyond) $z \sim 2$~\citep{pointecouteau2013,cucchetti2018,mernier2020}. Figure~\ref{fig:athena} illustrates this goal, by showing \FM{in the upper panel} a spectrum of a $z \sim 1$ cluster seen by different instruments, including \VB{the prediction for} \textit{Athena}/X-IFU. \FM{The lower panel shows how, even at $z=1$, key X/Fe are expected to be accurately measured \FMbis{when} assuming specific yields from AGB stars, SNcc and SNIa on a simulated cluster.} One can easily see that not only Fe, but \FM{the whole chemical composition of the ICM} will be constrained with exquisite  
\VB{accuracy} even at epochs where the Universe was half of its current age.

\FMbis{Unlike the near-Solar chemical composition we actually observe in the ICM (Sect.~\ref{sec:when:ratios}), the input yields proposed on the bottom panel of Fig.~\ref{fig:athena} show important discrepancies with the solar ratios. Such an exercise can thus be revised with more realistic yields that will be obtained from improved nucleosynthesis calculations (see below). In fact, simulation results are still largely affected  
by the uncertainties on the stellar yields and the adopted IMF. These refinements will be crucial for increasingly detailed comparisons with future observational findings, together with the generation of proper mock X-ray observations from simulated clusters, that are able to mimic the specifics of real X-ray observations and can allow for a more faithful comparison.}
\VBbis{Still from the theoretical point of view,}
improvements on the stellar evolution models themselves are needed in order to have more reliable estimates of chemical abundances in terms of absolute values.
As for the numerical modelling, simulations of galaxies and of galaxy clusters have reached an advanced level of detail and succeeded in matching many observational properties of both populations.
Nonetheless, higher numerical resolution and improvements in the sub-grid physical processes driving the baryonic evolution are still required. The ultimate goal remains indeed to match observed gas and stellar properties in clusters, namely the ICM and the cluster galaxy population, \textit{simultaneously}.
Higher resolution is necessary to simulate more realistic member galaxies, in addition to the central BCG, as they interact with the ambient ICM and contribute through several physical processes to the cluster enrichment with chemical elements produced within them, especially at early times.
The quest for higher resolution must of course be accompanied by a continuous refinement of the \VB{modelling of the} baryonic physical processes that are treated with sub-grid approaches and are often critically sensitive to numerical resolution.
Among the physical processes that play an important role in the production and distribution of chemical elements the treatment of dust is certainly a crucial, albeit still largely overlooked, one. This is of great importance for the properties of the galaxies and for the level of ICM enrichment, impacting both gas cooling and metal depletion. Besides early investigations with post-processing approaches, only recently a first self-consistent modelling of dust within a cosmological chemo-hydrodynamical simulation of galaxy clusters has been carried out~\cite[][]{gjergo2018}, finding promising results on the dust abundances and metallicity at the galaxy scale despite some discrepancy with the observed amount of dust in clusters, at low redshifts.

Although many questions are still open, the bright future offered by the next generation of X-ray observatories -- in synergy with constantly improved cosmological simulations, stellar nucleosynthesis yields, and spectral codes -- 
will allow us to ultimately unveil the details of the cosmic cycle of metals in our Universe.

\bibliography{author}{}
\bibliographystyle{spbasic}



\end{document}